\documentclass[a4paper,11pt]{article}
\pdfoutput=1 

\usepackage{jcappub} 
\usepackage[T1]{fontenc} 

\usepackage{xcolor}

\def\lsim{\mathrel{\raise.3ex\hbox{$<$\kern-.75em\lower1ex\hbox{$\sim$}}}}
\def\gsim{\mathrel{\raise.3ex\hbox{$>$\kern-.75em\lower1ex\hbox{$\sim$}}}}

\title{Searching For Superheavy Decaying Particles With Ultra-High-Energy Neutrino Observatories}

\author[a]{Kim V.~Berghaus,}
\author[b,c,d,e,f]{Dan Hooper,}
\author[d]{Emily R.~Simon}

\affiliation[a]{Walter Burke Institute for Theoretical Physics, California Institute of Technology}
\affiliation[b]{Wisconsin IceCube Particle Astrophysics Center, University of Wisconsin-Madison}
\affiliation[c]{Department of Physics, University of Wisconsin-Madison}
\affiliation[d]{Department of Astronomy \& Astrophysics, University of Chicago}
\affiliation[e]{Kavli Institute for Cosmological Physics, University of Chicago}
\affiliation[f]{Theoretical Astrophysics Department, Fermi National Accelerator Laboratory}

\emailAdd{berghaus@caltech.edu}
\emailAdd{dwhooper@wisc.edu}
\emailAdd{ersimon@uchicago.edu}

\abstract{If there exist unstable but long-lived relics of the early universe, their decays could produce detectable fluxes of gamma rays and neutrinos. In this paper, we point out that the decays of superheavy particles, $m_{\chi} \gtrsim 10^{10} \, \text{GeV}$,
would produce an
enhanced flux of ultra-high-energy neutrinos through the processes of muon and pion pair production in the resulting electromagnetic cascades. These processes transfer energy from electromagnetic decay products into neutrinos, relaxing the constraints that can be derived from gamma-ray observations, and increasing the sensitivity of high-energy neutrino telescopes to superheavy particle decays. Taking this into account, we derive new constraints on long-lived superheavy relics from the IceCube Neutrino Observatory, and from the Fermi Gamma-Ray Space Telescope. We find that IceCube-Gen2, and other next generation neutrino telescopes, will provide unprecedented sensitivity to the decays of superheavy dark matter particles and other long-lived relics. }

\begin{document}
\maketitle
\flushbottom

\section{Introduction}

Although the dark matter of our universe must be relatively stable, it is possible that the particles that make up this substance could decay on timescales much longer than the age of the universe~\cite{Poulin:2016nat}. On the other hand, unstable relics that make up only a small fraction of the dark matter could decay on much shorter timescales. Constraints have been placed on decaying dark matter and other unstable relics using a variety of astrophysical messengers, including gamma-rays~\cite{Das:2023wtk,LHAASO:2022yxw,Blanco:2018esa,Cohen:2016uyg,Ando:2015qda,Hutsi:2010ai,Murase:2015gea,Murase:2012xs,Fermi-LAT:2015kyq,2012ApJ...761...91A,Kalashev:2016cre,Cirelli:2012ut,Esmaili:2015xpa,Liu:2016ngs}, X-rays~\cite{Boyarsky:2007ge,Yuksel:2007xh,Perez:2016tcq}, neutrinos~\cite{Das:2024bed,Guepin:2021ljb,Murase:2012xs,Palomares-Ruiz:2007egs}, and cosmic rays~\cite{Nguyen:2024kwy,Das:2023wtk,Ibarra:2013zia}.

In this study, we revisit the flux of ultra-high-energy neutrinos that would be produced through the decays of dark matter particles or other superheavy relics. We point out here that photons injected at redshift, $z$, with energy, $E_{\gamma } \gsim 10^{10} \, {\rm GeV} \, (1+z)^{-1}$, can scatter with the cosmic microwave background (CMB) to produce pairs of muons or charged pions, which go on to generate neutrinos through their decays. At extremely high energies, these processes can effectively transfer an order one fraction of the total energy in photons and electrons into ultra-high-energy neutrinos, suppressing the constraints on superheavy particle decay that can be derived from gamma-ray observations, and creating an opportunity to probe such scenarios with ultra-high-energy neutrino observatories.  

Previous studies have found unstable particle species to be most strongly constrained by gamma-ray observations.
After taking into account the contributions from muon and pion pair production, however, we find that superheavy particles with lifetimes in the range of $\tau_{\chi} \sim 10^{16}-10^{17} \, {\rm s}$ are already more strongly constrained by the IceCube Neutrino Observatory than by the Fermi Gamma-Ray Space Telescope. Future high-energy neutrino telescopes, such as IceCube-Gen2~\cite{IceCube-Gen2:2020qha}, are expected to improve upon these limits by another two orders of magnitude. For unstable particles with lifetimes much longer than the age of the universe, we find that IceCube-Gen2 will provide the greatest sensitivity in the mass range of $m_\chi \gsim 10^{13} - 10^{16} \, {\rm GeV}$, depending on the decay channel.


\section{Theoretical Motivation}

If a long-lived particle species heavier than $\mathcal{O}(100 \, {\rm TeV})$ were to reach equilibrium with the Standard Model bath in the early universe, the energy density of these particles would ultimately come to exceed the measured density of dark matter~\cite{Griest:1989wd}. This has long motivated the consideration of lighter dark matter candidates, such as those with weak-scale masses. If the dark matter is very feebly interacting, however, equilibrium may have never been achieved, opening a window for one to consider very heavy -- and even superheavy -- relics from the early universe.

Gravitational particle production is perhaps the most well-motivated means by which a population of superheavy particles may have been generated in the early universe~\cite{Chung:1998zb} (for a review, see Ref.~\cite{Kolb:2023ydq}). This mechanism is typically most effective for particles with masses near the energy scale of inflation, $m_{\chi} \sim H_I$, and could be responsible for generating all or some of our universe's dark matter. Alternatively, superheavy particles may have been produced in the early universe through the process of thermal freeze-in~\cite{Kolb:2017jvz}, or other proposed mechanisms~\cite{Babichev:2018mtd,Kim:2019udq,Dudas:2020sbq}.

If a superheavy particle species is able to decay through anything other than very high-dimensional operators, its lifetime will be much shorter than the age of the universe. Parametrically, we expect a Planck-suppressed, dimension-$D$ operator to lead to lifetimes on the order of $\tau_{\chi} \sim M_{\rm Pl}^{2D-8}/m^{2D-7}_{\chi}$. To obtain a cosmologically long-lived lifetime thus requires all lower-dimensional ($D \le 5 $) operators to be forbidden. Some ideas in this direction have involved accidental symmetries in a secluded sector~\cite{Kolb:1998ki}, or decays that proceed only through non-perturbative mechanisms~\cite{Kuzmin:1997jua}. Additionally, the superheavy relics in question could be dark blobs or dark quark nuggets, for which stability could be ensured if the self-energy of the composite object is smaller than the sum of the individual constituents~\cite{Grabowska:2018lnd,Bai:2018dxf}.

\section{The Initial Spectrum of Superheavy Particle Decay Products}

We begin by quantifying the prompt spectra of the neutrinos and gamma rays that result from the decays of superheavy dark matter or other long-lived relics. In doing so, we consider the following representative decay channels:\footnote{Other hadronic decay channels will result in spectra that are very similar to that found in the case of $\chi \rightarrow b\bar{b}$.} 
\begin{equation}
\label{eq:channel}
\chi \to b\bar{b}, \, v_e \bar{v}_e, \, e^+e^-, \, \mu^+ \mu^-, \, \tau^+\tau^-, \,W^+W^-, \,ZZ, \,\gamma \gamma.
\end{equation}

\begin{figure}
    \centering
    \includegraphics[width=0.49\textwidth]{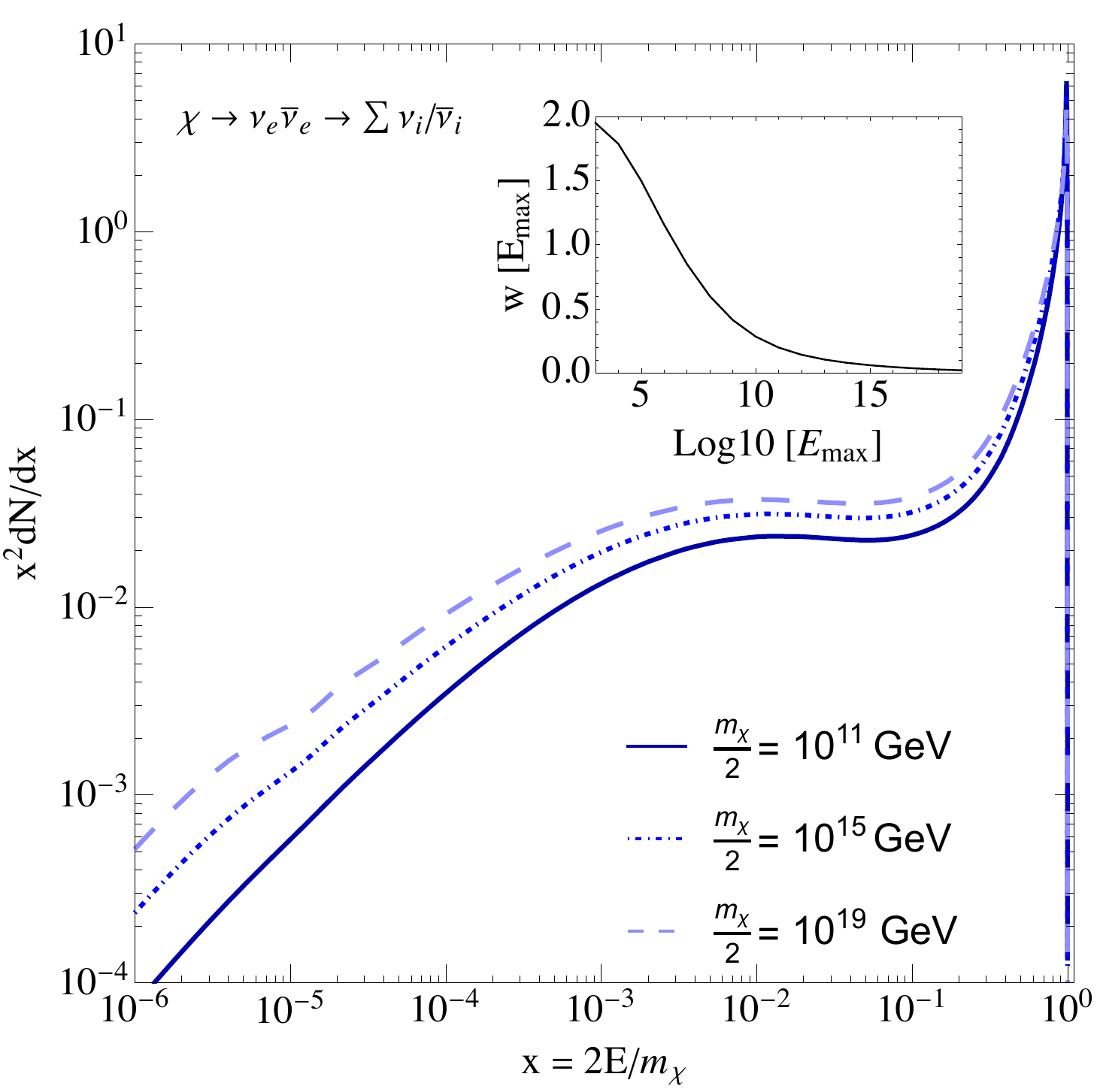}
    \includegraphics[width=0.49\textwidth]{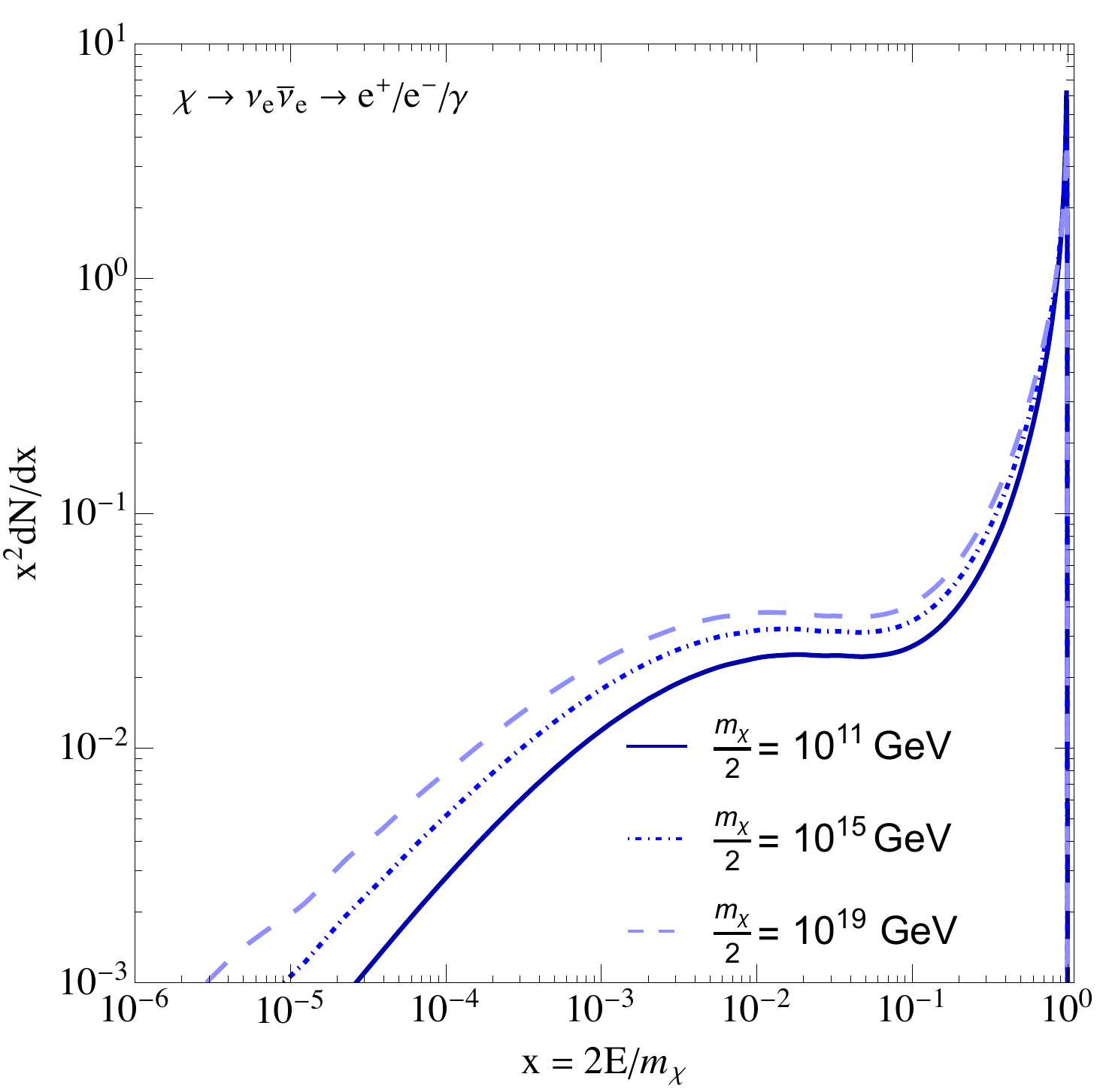}
    \caption{The prompt, all-flavor spectrum of neutrinos and antineutrinos (left), and of photons, electrons, and positrons (right), for superheavy particles decaying to electron neutrinos, $\chi \to \nu_e \bar{\nu}_e$, including electroweak corrections. These results were obtained using the publicly available code, HDMSpectra~\cite{Bauer:2020jay}. The coefficient, $w(E_{\text{max}})$, as shown in the left inset, illustrates the redistribution of energy away from the two hard final states and is defined in Eq.~\ref{eq:w}.}
    \label{fig:promptspectra}
\end{figure}

\begin{figure}
    \centering
    \includegraphics[width=0.49\textwidth]{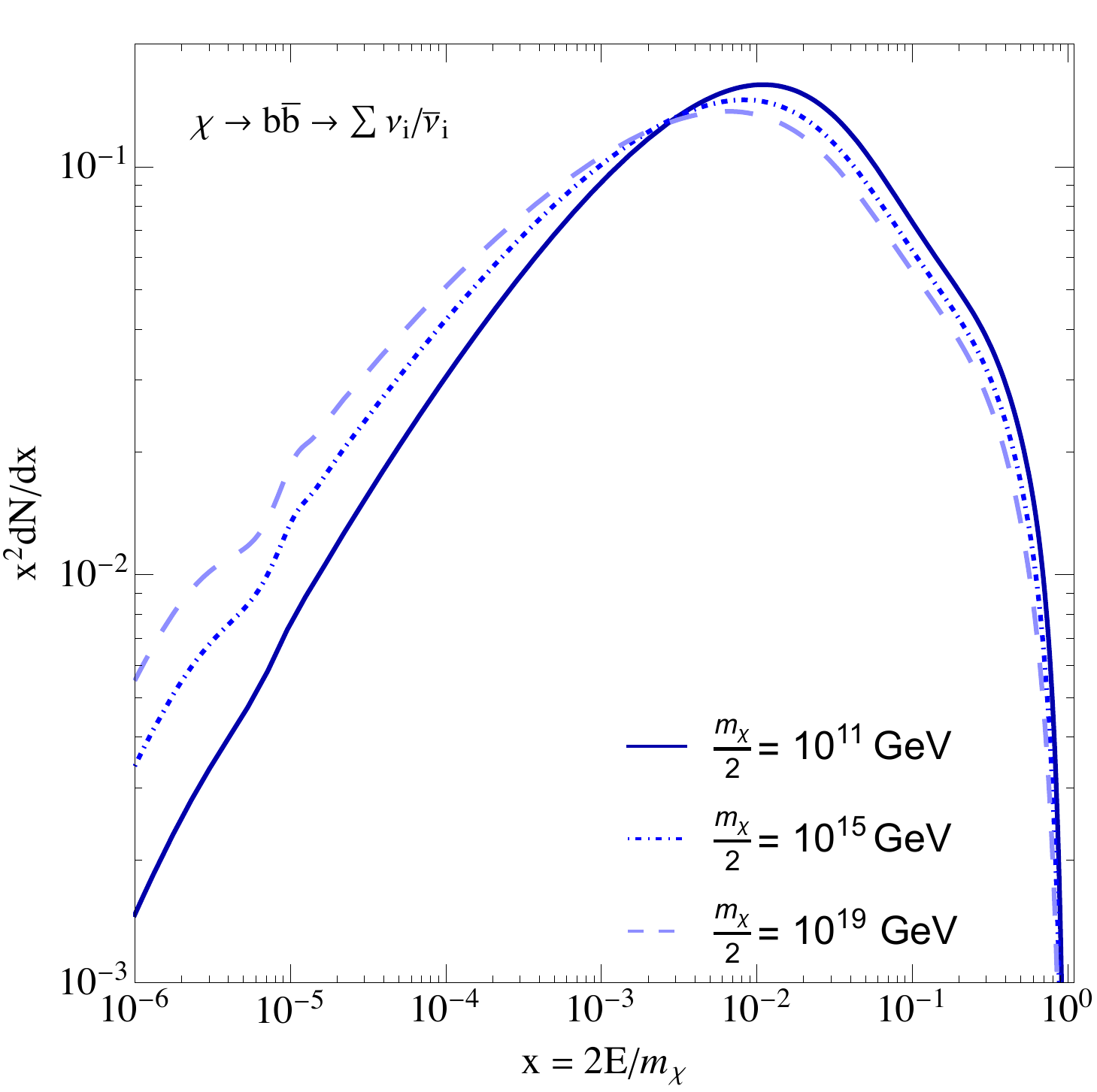}
    \includegraphics[width=0.49\textwidth]{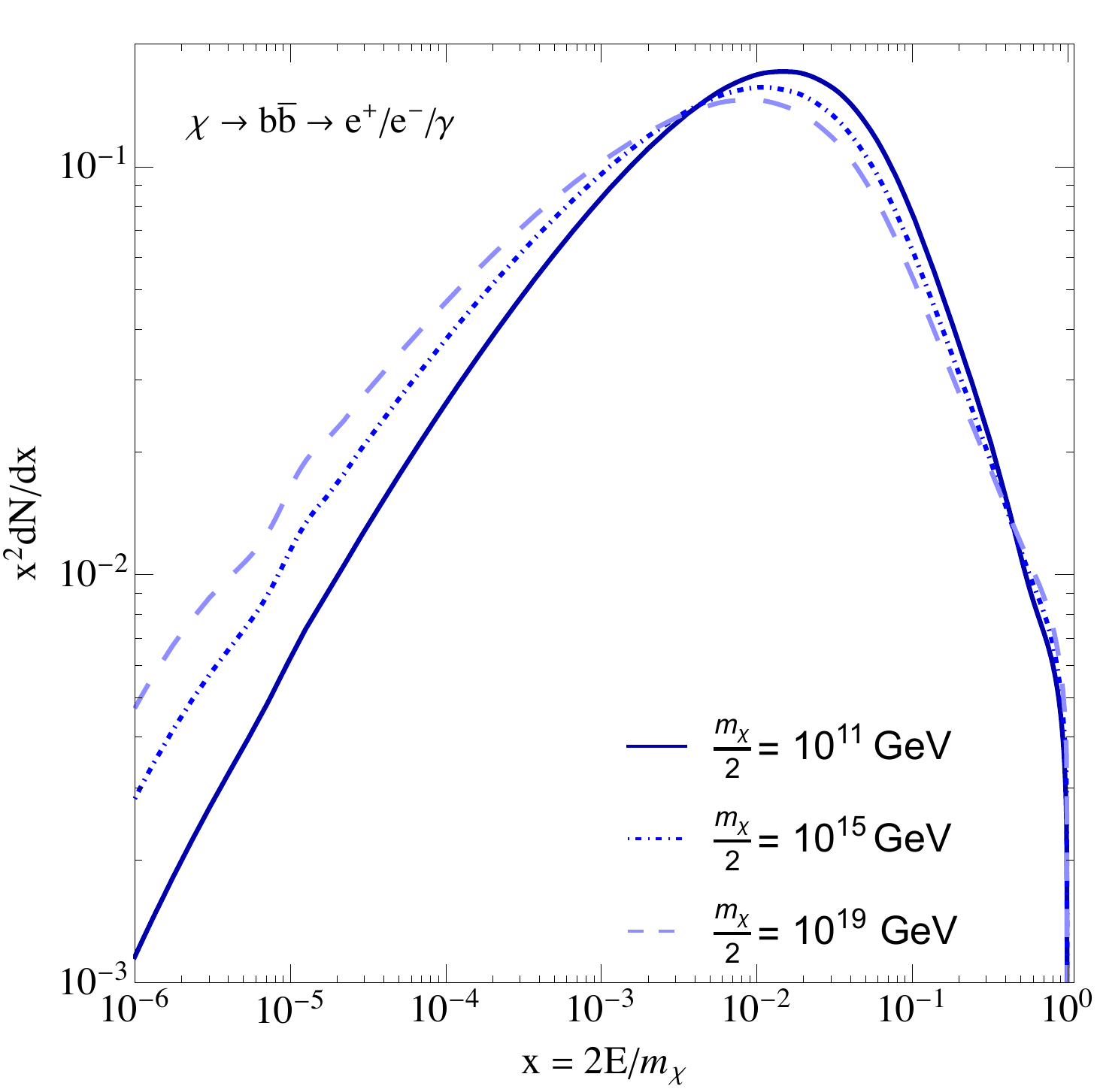}
    \caption{As in Figure \ref{fig:promptspectra}, but for the decay channel, $\chi \to b\bar{b}$.}
    \label{fig:promptspectrabb}
\end{figure}

For each of these primary decay channels
we determine the prompt, all-flavor neutrino flux, $ \sum_i  dN_{\nu_i}/dE$.
We include electroweak corrections in our calculations, which increase with the value of the decaying particle's mass, $m_\chi$. For $m_{\chi} \gg {\rm TeV}$, electroweak gauge bosons are effectively massless and are radiated with probabilities exceeding one due to large Sudakov logarithms that scale as $\propto \alpha_w/4\pi \log(m^2_{\chi}/m^2_V)$, overcoming the phase-space suppression from higher-multiplicity final states.
To determine the number and energy distribution of the neutrinos in the final state requires us to resum all of the splittings which one can achieve via electroweak showering~\cite{Chen:2016wkt, Berghaus:2018zso}, or by evolving the full set of electroweak DGLAP equations~\cite{ALTARELLI1977298, Dokshitzer:1977sg, Gribov:1972ri,
Kalashev:2016cre, Bauer:2020jay}.
Qualitatively, electroweak corrections smear out the delta function of the primary decay products. 
For example, in the absence of electroweak corrections, the neutrino spectrum associated with the primary channel, $\chi \to \nu_e \bar{\nu}_e $, is given by
\begin{equation}
\label{eq:delta}
\frac{dN_{\nu_e}}{dE} + \frac{dN_{\bar{\nu}_e}}{dE}
= 2 \delta\left(\frac{m_{\chi}}{2} - E\right) \, ,
\end{equation}
resulting in one electron neutrino and one anti-electron neutrino, each with $E_{\nu_e} = E_{\bar{\nu}_e} = m_{\chi}/2$. Electroweak corrections shift the resulting distribution from two particles with equal energies to many particles with lower energies. This decreases the weight of the delta function coefficient appearing in Eq.~\ref{eq:delta} from $2$ to a smaller value, $w(E)$, as shown in the left inset of Figure~\ref{fig:promptspectra}.
In Figure \ref{fig:promptspectra}, we have summed over flavors, and over particles and antiparticles, such that the total neutrino flux can be written as
\begin{align}
\label{eq:w}
\frac{dN}{dx} (x, E_{\text{max}}) & = \sum_i \frac{dN_{\nu_i}}{dx} (x,E_{\text{max}}) + \frac{dN_{\bar{\nu}_i}}{dx} (x,E_{\text{max}})  \nonumber\\
 & = w(E_\text{max}) \delta(1-x) + 2 \sum_i \frac{dN_{\nu_i}}{dx}(x,E_{\text{max}})~,
\end{align}
where $E_{\rm max} = m_{\chi}/2$. 

All of the particles that are produced in these decays eventually decay into stable species, including photons, neutrinos, electrons, protons, and their respective antiparticles. Electroweak corrections ensure that an $\mathcal{O}$(1) fraction of the total energy goes into neutrinos, regardless of the decay channel. Similarly, even for a channel such as $\chi \rightarrow \nu_e \bar{\nu}_e$, electroweak corrections guarantee that an 
$\mathcal{O}$(1) fraction of the total energy is carried away by electromagnetically interacting particles. 
This is illustrated in Figure~\ref{fig:promptspectra}, where we show in the right panel the prompt decay spectrum summed over electrons, positrons and photons for a selection of masses. 
For comparison, we show the corresponding spectrum for the hadronic $\chi \to b\bar{b}$ decay channel in Figure~\ref{fig:promptspectrabb}. To calculate these spectra, we have used the publicly available code, HDMSpectra~\cite{Bauer:2020jay}.

The impact of electroweak corrections on the observational signatures of superheavy particle decay has been considered in previous studies~\cite{Cohen:2016uyg, Berghaus:2018zso, Das:2024bed}. The neutrino flux that is generated in the propagation of the electromagnetic decay products through the processes of muon pair production and pion pair production has, however, not been quantified before. For sufficiently massive decaying particles, these processes qualitatively change the resulting constraints and prospects for detection. 


\section{Neutrinos from Muon and Pion Pair Production}
\label{sec:muons}

Ultra-high-energy photons can scatter with the cosmic microwave background (CMB) through the Breit-Wheeler process~\cite{Breit:1934} to produce an electron-positron pair. At even higher energies, it becomes possible for a photon to pair produce other species of charged particles, including muons, tau leptons, and charged pions, which decay to produce neutrinos~\cite{Hooper:2023ssc}. This requires the energy in the center-of-momentum frame to exceed the total mass of the produced particles, $2 m_{\mu, \pi, \tau} < E_{\rm CM} = [2E_\gamma \epsilon \,(1-\cos\theta)]^{1/2}$, where $E_\gamma$ is the energy of ultra-high-energy photon, $\epsilon$ is the energy of the target CMB photon, and $\theta$ is the angle between them. Due to the very high energy thresholds for these processes, they are generally restricted to extremely high-energy photons, or to photons produced at very high redshifts.

\begin{figure}
    \centering
    \includegraphics[width=0.55\textwidth]{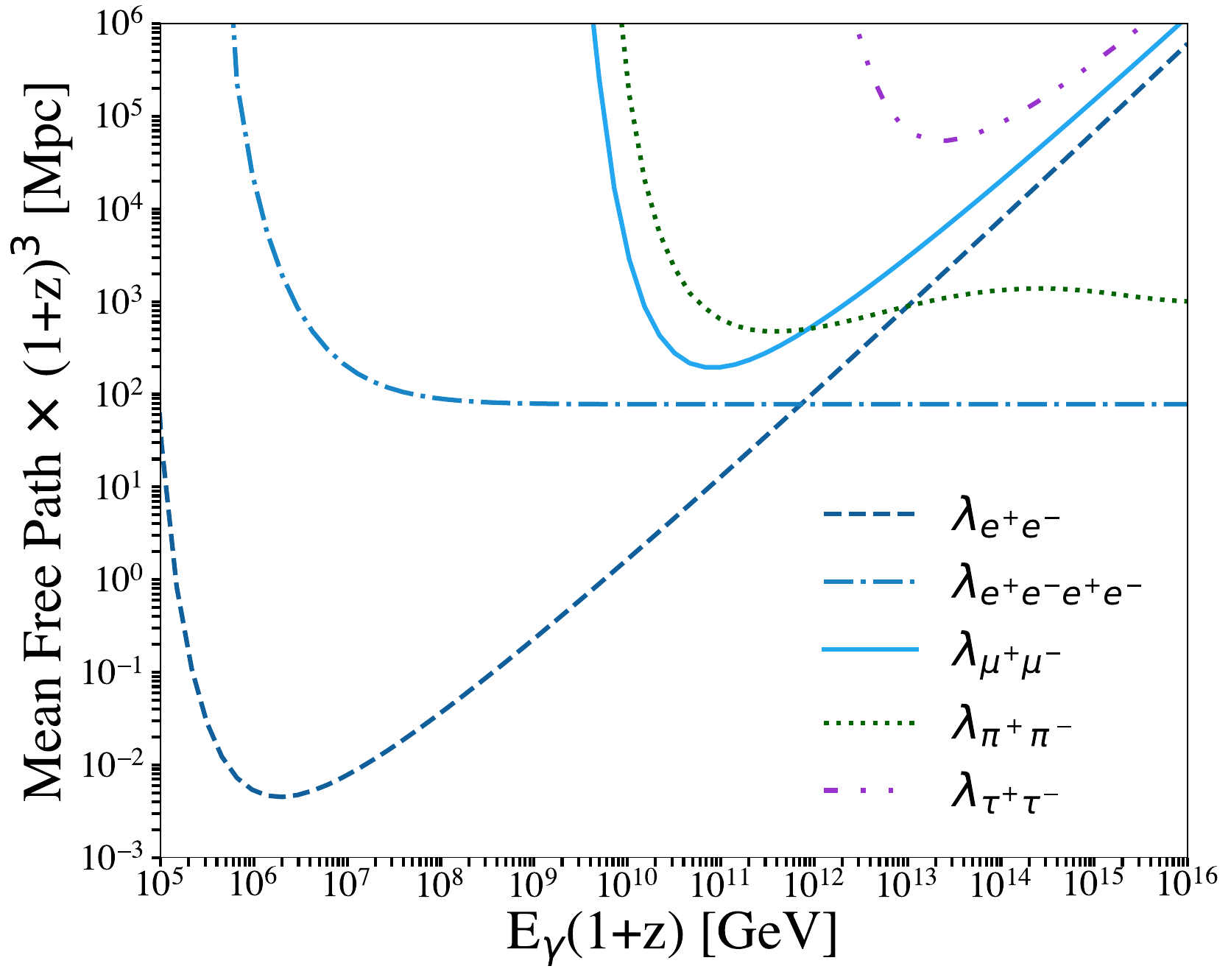}
    \caption{The mean free path for an ultra-high-energy photon propagating through the CMB for each of the processes considered in this study.}
\label{fig:characteristic_length}
\end{figure}

The cross section for charged lepton pair production is given by~\cite{Akhiezer:1965}
\begin{equation}
    \sigma_{\gamma \gamma \rightarrow l^+ l^-} = \frac {2\pi \alpha^2}{E^2_{\rm CM}} \left[2\beta(\beta^2 -2)+(3-\beta^4) \ln\left(\frac{1+\beta}{1-\beta} \right) \right],
\end{equation}
where
\begin{equation}
    \beta = \left[1-\left(\frac{2m_{e,\mu,\tau}}{E_{\rm CM}} \right)^2 \right]^{1/2},
\end{equation}
and $\alpha$ is the fine structure constant.
Photons that exceed the threshold for muon pair production will also be capable of undergoing double pair production, $\gamma + \gamma \rightarrow e^+ + e^- + e^+ + e^-$, with a cross section of
\begin{equation}
    \sigma_{\gamma \gamma \rightarrow e^+ e^- e^+ e^-} \approx \bigg( 1- \frac{4m_e^2}{E_\gamma \epsilon}  \bigg)^6 \, \sigma_{\gamma \gamma \rightarrow e^+ e^- e^+ e^-}^{\infty} \, ,
\end{equation}
where $\sigma_{\gamma \gamma \rightarrow e^+ e^- e^+ e^-}^{\infty} \approx 6.45 \times 10^{-30}$ cm$^2$~\cite{Brown:1973onk}. 

Processes producing hadrons can be relevant if $\sqrt{s} \geq 2 m_{\pi^{\pm}}$, with charged pions being the lightest and thus easiest to produce. 
The cross section for charged pion pair production has been measured at colliders up to energies of $ \sqrt{s} \sim 10^{2}$ GeV~\cite{Godbole:2003wj, Galanti:2019rnl}.
Beyond this, we conservatively estimate that this cross section remains constant, at $\sigma_{\pi^+ \pi^-} \approx 5 \times 10^{-31}$ cm$^2$, likely underestimating the true cross section at the highest energies. In Figure~\ref{fig:characteristic_length}, we plot the mean free paths for each of these process as a function of photon energy (see also, Ref.~\cite{Esmaeili:2022}). Comparing the mean free paths shown in Figure~\ref{fig:characteristic_length} to the Hubble length, $c/H_0 \approx 4280 \, {\rm Mpc}$, we conclude that the cascade will evolve quickly relative to the timescales associated with cosmological expansion, allowing us to treat the evolution of the cascade as instantaneous in our calculations.

To evaluate the impact of these processes, we have written a Monte Carlo to simulate the propagation of particles through the CMB. We allow each ultra-high-energy photon to produce electrons, muons, pions, taus, and their antiparticles, according to the mean free path for each process. We then calculate the spectrum of the resulting decay products. Any electrons and positrons that are produced in this way undergo inverse Compton scattering with CMB photons in the Klein-Nishina limit, resulting in the production of new photons with $E^{'}_{\gamma} \approx E_e \approx E_{\gamma}/2$. This cycle repeats for any new photons that are produced, resulting in an electromagnetic cascade that evolves until none of its particles have enough energy to produce any further muon or pion pairs. Once we have the resulting spectra of neutrinos and photons in this cascade, we calculate the flux of these particles that reach Earth, including the effects of cosmological redshift (see Sec.~\ref{sec:DM}).

\begin{figure}
    \centering
    \includegraphics[trim = 5 0 0 0, clip=true, width=0.325\textwidth]{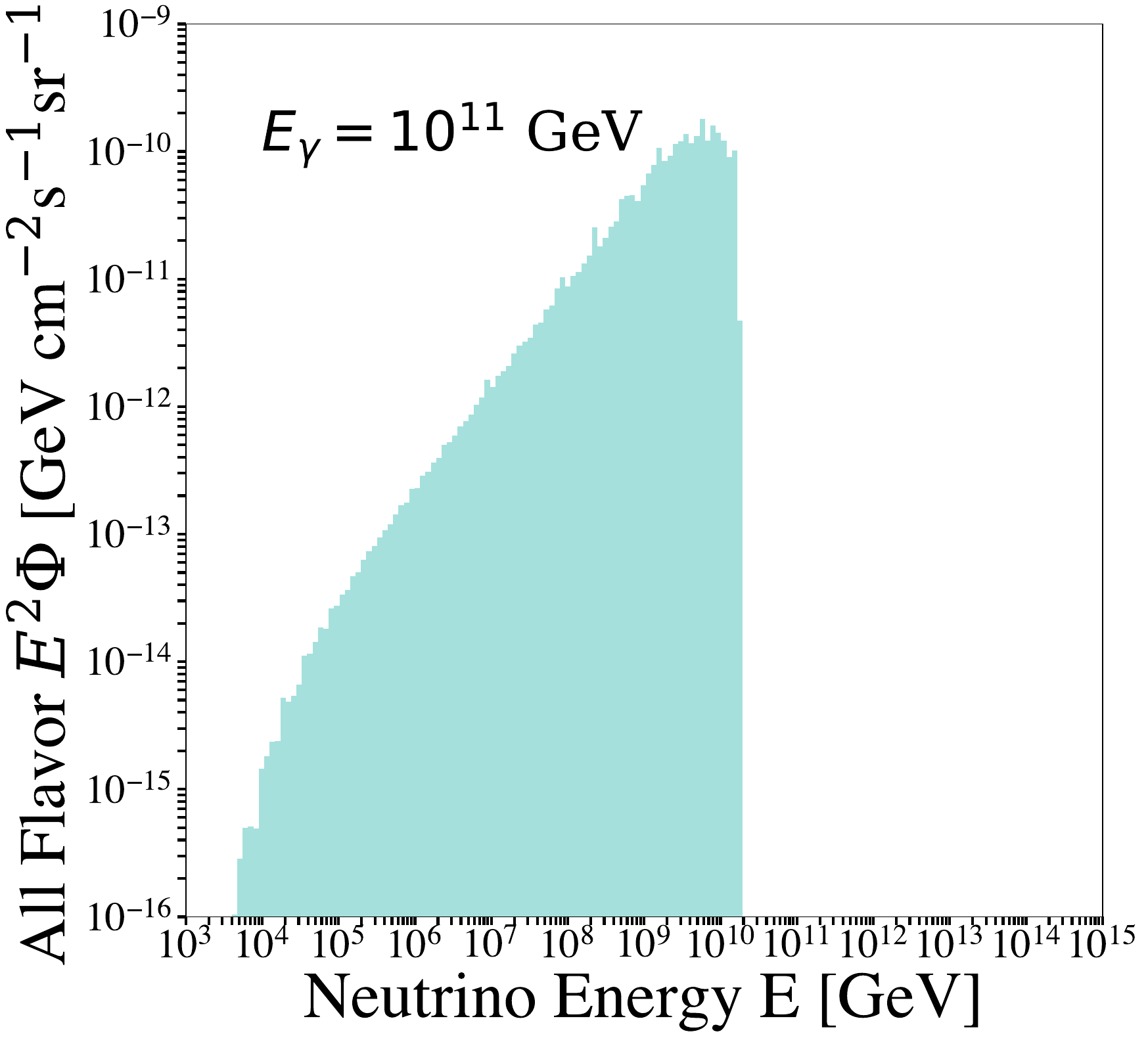}
    \includegraphics[trim = 5 0 0 0, clip=true, width=0.325\textwidth]{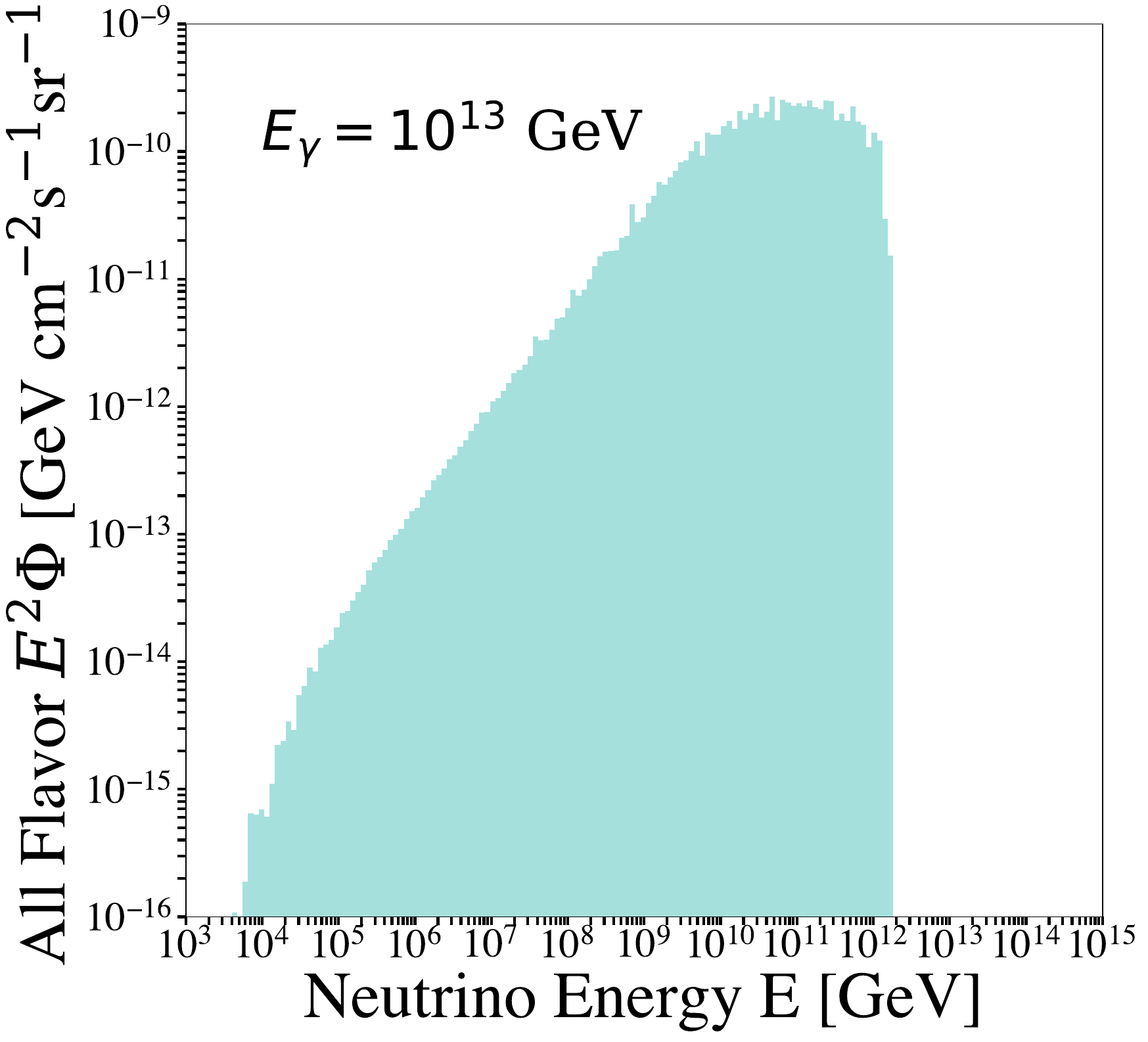}
    \includegraphics[trim = 5 0 0 0, clip=true,
    width=0.325\textwidth]{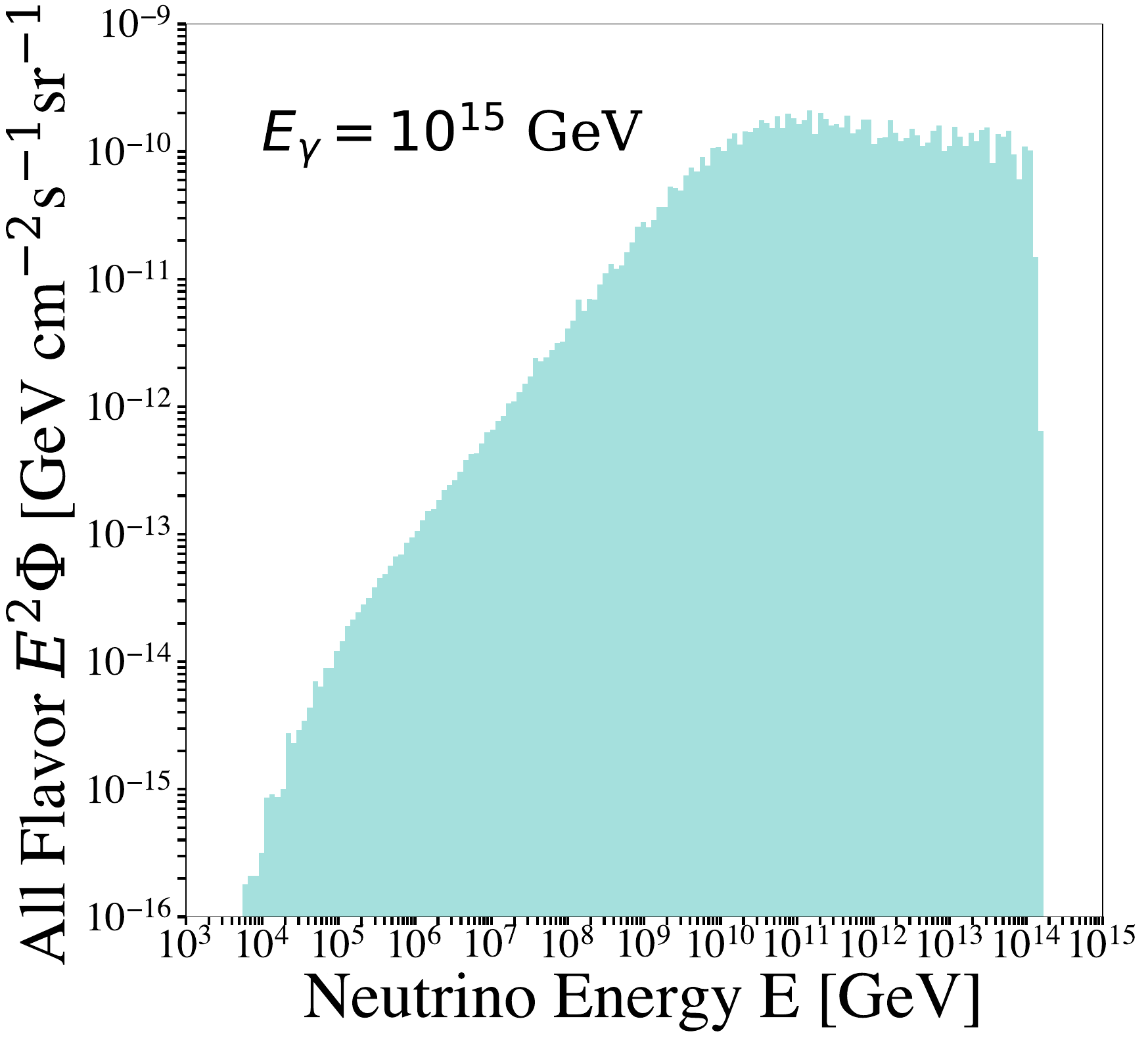}
    \caption{The spectrum of neutrinos generated through muon, pion, and tau pair production, for the case of a mono-energetic spectrum of ultra-high-energy photons injected as the decay products of a dark matter particle with a lifetime of $\tau_{\chi} = 5 \times 10^{29} \,{\rm s}$.  Results are shown for primaries with $E_{\gamma}=10^{11}$, $10^{13}$, and $10^{15} \, {\rm GeV}$.}
    \label{fig:neutrinos}
\end{figure}

\begin{figure}
    \centering
    \includegraphics[width=0.55\textwidth]{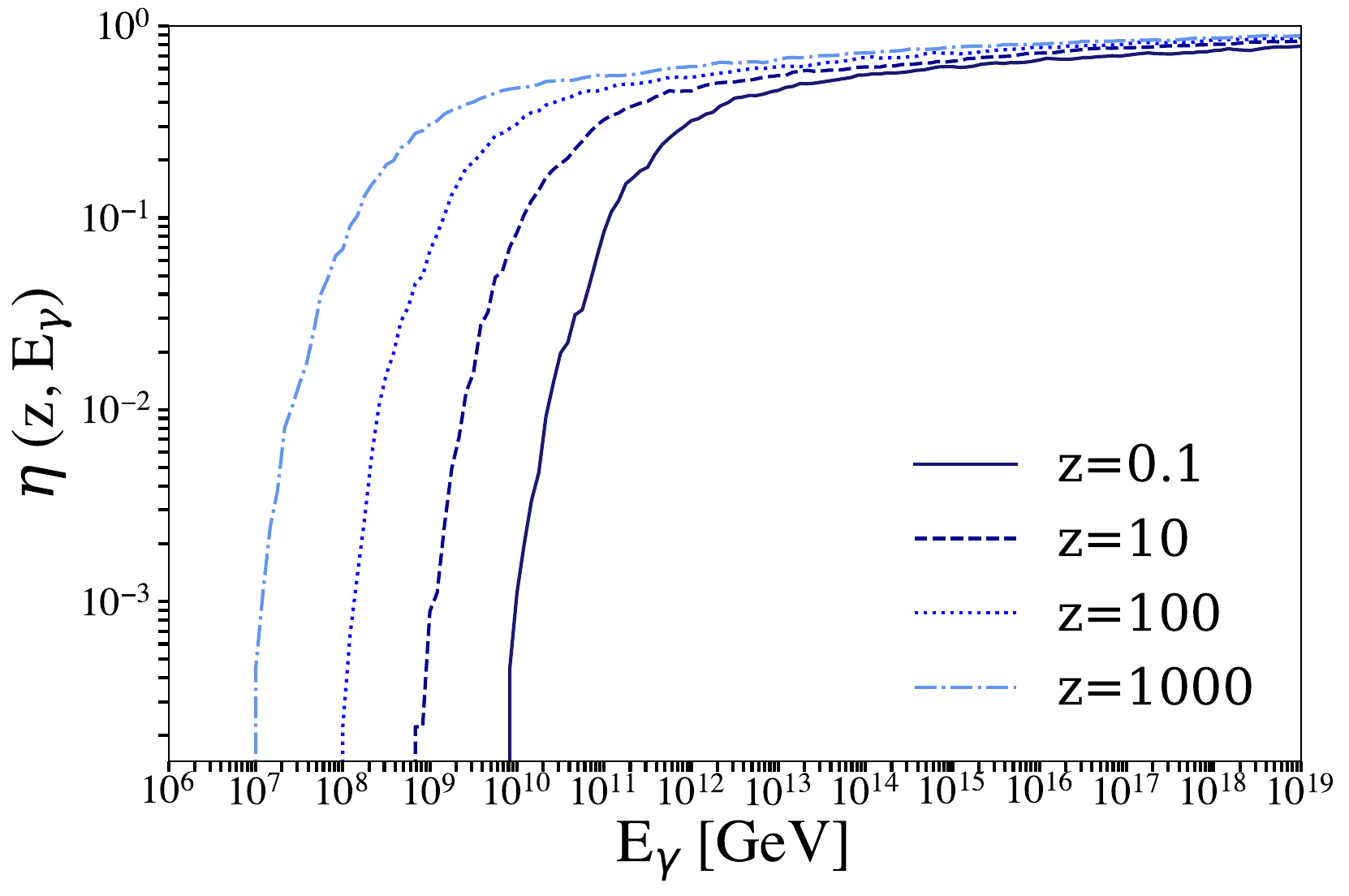}
    \caption{The average fraction of an ultra-high-energy photon's energy that goes into neutrinos as it propagates through the CMB, $\eta(z, E_{\gamma})$. The remaining fraction of this energy goes into the production of an electromagnetic cascade. }
    \label{fig:nu_efficiency}
\end{figure}

In Figure~\ref{fig:neutrinos}, we plot the spectrum of neutrinos that results from muon, pion, and tau pair production, for a case of a mono-energetic spectrum of ultra-high-energy photons injected as the decay products of a dark matter particle with a lifetime of $\tau_{\chi} = 5 \times 10^{29} \,{\rm s}$. For sufficiently energetic photons, these processes can transfer an order one fraction of their initial energy into ultra-high-energy neutrinos. This is illustrated in Figure~\ref{fig:nu_efficiency}, where we plot the average fraction of an ultra-high-energy photon's energy that is transferred into neutrinos as it propagates through the CMB. This result might seem surprising given that, at all energies, the mean free paths for pair production and/or double pair production are shorter than those associated with muon, pion, or tau pair production. While this is true, each time that a photon undergoes electron-positron pair production (or double pair production), new photons are produced which then get another opportunity to undergo muon, pion, or tau pair production. In contrast, once neutrinos are produced through muon or pion decay, those particles remain in that state until they reach Earth.\footnote{At $z=0$, the mean free path for ultra-high-energy neutrino scattering with the cosmic neutrino background exceeds $\sim 40 \, {\rm Gpc}$ for all values of $E_{\nu}$. At significantly greater redshifts, however, the larger number density and temperature of the background neutrinos can cause this process to be potentially important for neutrinos with enough energy to scatter through the $Z$ resonance~\cite{Ema:2013nda, Das:2024bed}.}

\begin{figure}
    \centering
    \includegraphics[width=0.49\textwidth]{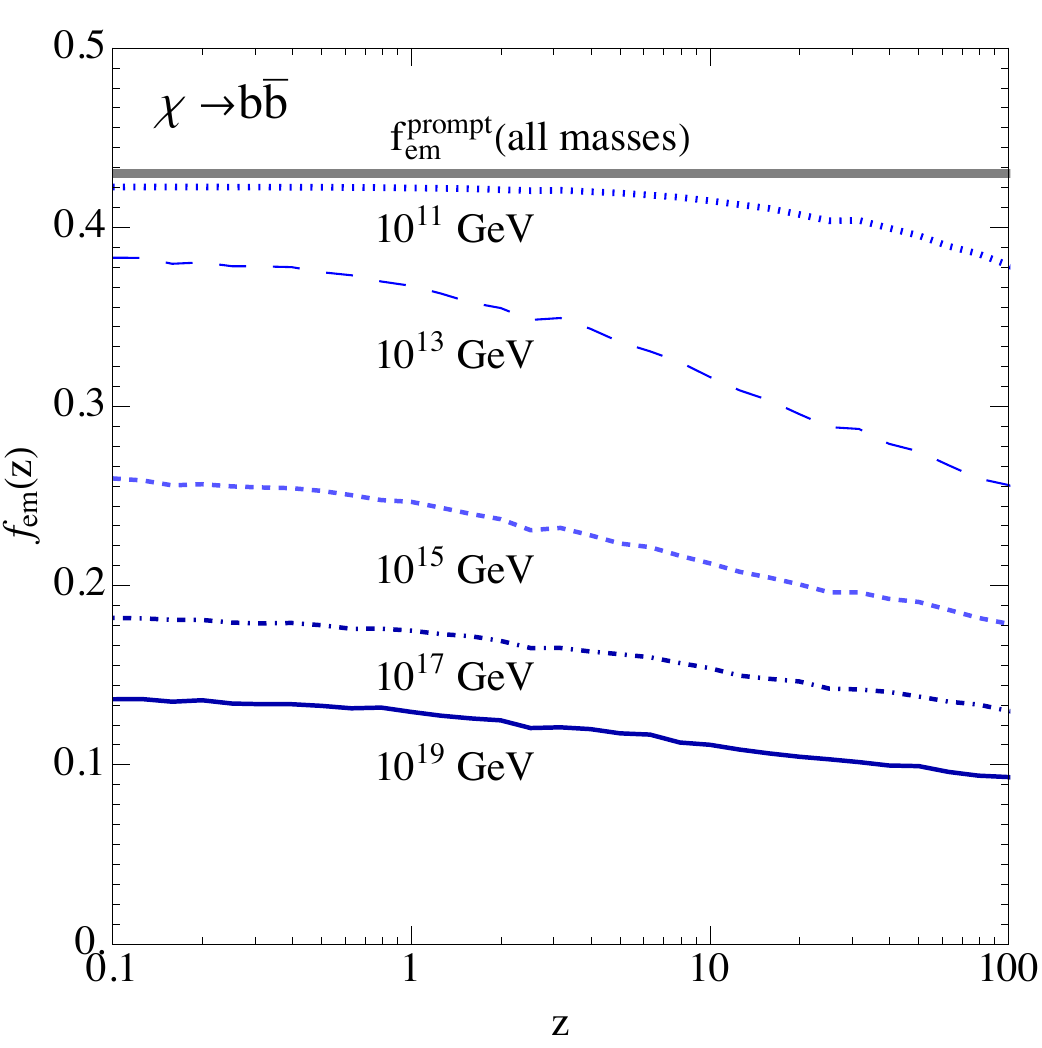}
    \includegraphics[width=0.49\textwidth]{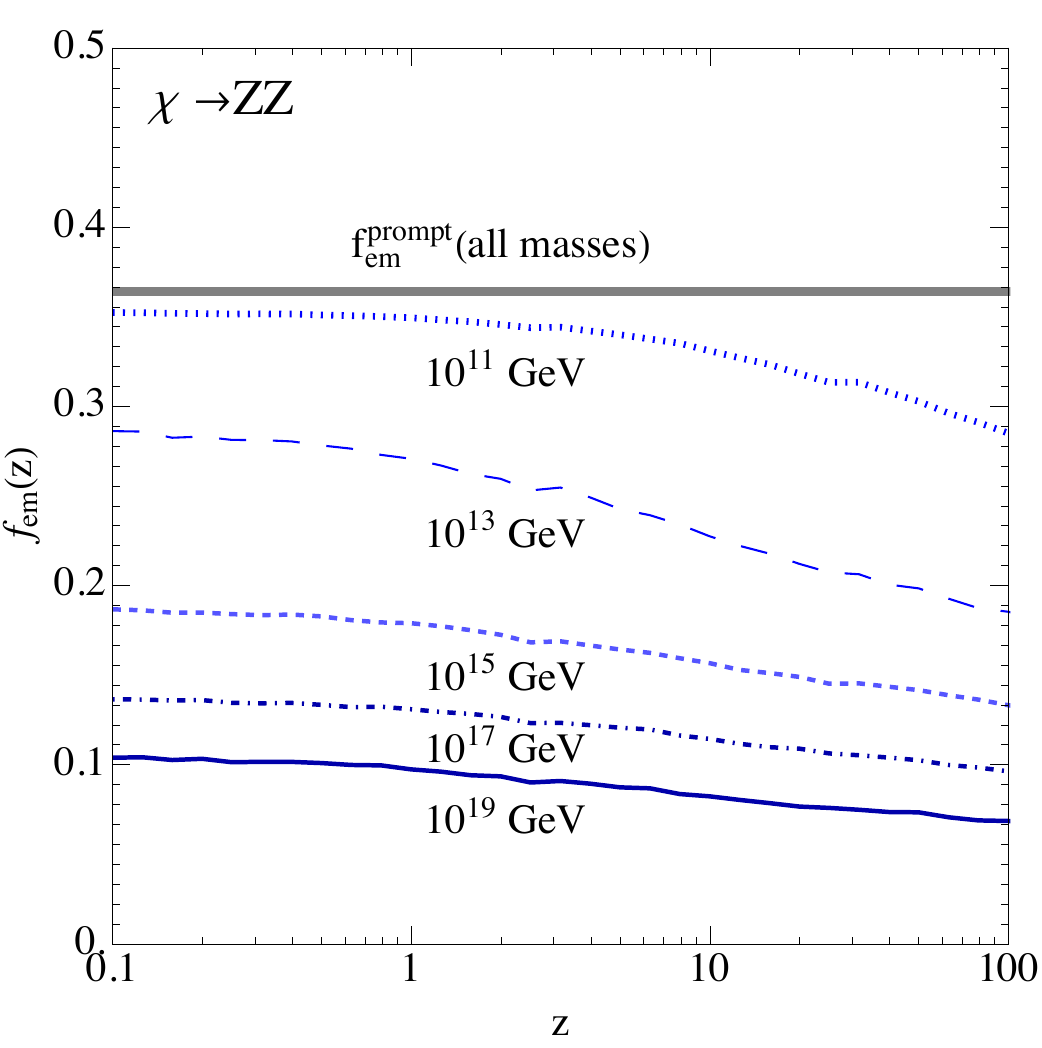}
    \caption{The average fraction of the total energy from a particle decay at redshift, $z$, that goes into the formation of an electromagnetic cascade. Results are shown for the decay channels $\chi \rightarrow b\bar{b}$ (left) and $\chi \rightarrow ZZ$ (right), and for several values of the decaying particle's mass, between $m_{\chi}=10^{11} \, {\rm GeV}$ and
 $10^{19} \, {\rm GeV}$. 
    }
    \label{fig:fem}
\end{figure}

For particles decaying at a redshift, $z$, the fraction of the total energy that ultimately goes into photons, electrons, or positrons (and, therefore, into the electromagnetic cascade) is given by
\begin{equation}
f_{\text{em}} (z) =\frac{1}{2 E_{\text{max}}} \int_{E_{\text{min}}}^{E_{\text{max}}} E \frac{dN}{dE} \, \left(1-\eta(z, E) \right) dE\, ,
\end{equation}
where $dN/dE$ is the prompt spectrum of photons, electrons, and positrons, as shown in the right frames of Figures~\ref{fig:promptspectra} and~\ref{fig:promptspectrabb}, $\eta$ is the average fraction of the energy in a photon (or electron/positron) that goes into neutrinos, as shown in Figure~\ref{fig:nu_efficiency}, and $E_{\max} = m_{\chi}/2$. We plot this quantity in Figure~\ref{fig:fem} as a function of redshift, for the decay channels $\chi \rightarrow b\bar{b}$ and $\chi \rightarrow ZZ$, and for several values of $m_{\chi}$. We also show in this figure the fraction of the energy that is predicted from the prompt emission alone, neglecting the impact of the cascade (corresponding to the $\eta = 0$ limit). In that scenario, there is only a slight dependence on the mass, resulting from the electroweak corrections.

To calculate the contribution to the isotropic gamma-ray background from superheavy particle decays, we follow the derivation of the redshift-dependent shape of the photon spectrum due to electromagnetic energy injection, $\mathcal{L}(E,z)$, as presented in Refs.~\cite{Zsziarski:1989abc, Zziarski:1988}, and which was applied to the case of long-lived relics in Ref.~\cite{Berghaus:2018zso}. The derived spectral shape takes into account the relevant processes of the electromagnetic cascade that down-convert high energy photons until the universe becomes transparent to them~\cite{
Zsziarski:1989abc, Zziarski:1988, Blanco:2018bbf,Blanco:2018esa}. We normalize this spectral shape according to the total amount of electromagnetic energy that is injected, quantified by $f_{\chi}$ and $f_{\text{em}}(z)$, as shown in Figure \ref{fig:fem}. The resulting photon flux today is given by 
\begin{equation}
\label{eq:gammacon}
\Phi_{\gamma}(E) = \frac{\rho_{\text{crit}} f_{\chi} \Omega_\text{dm}  }{4 \pi \tau_{\chi}} \int  \frac{ f_{\text{em}}(z) \, \mathcal{L} (E(1+z),z) \,e^{-t(z)/\tau_{\chi}}}{H(z)}  \, dz ,
\end{equation} 
where $\rho_{\rm crit}\approx 4.94 \times 10^{-6} \text{GeV}/\text{cm}^3$, $\Omega_\text{dm} \approx 0.265$ and $H(z)$ is the Hubble rate. Throughout our calculations, we have adopted $\Omega_{\Lambda} = 0.69$, $\Omega_M = 0.31$, $\Omega_R = 9 \times 10^{-5}$, and $H_0 = 67.7 \text{km}/\text{s}/\text{Mpc}$~\cite{Planck:2018vyg}.
For further details regarding this procedure, see Ref.~\cite{Berghaus:2018zso}. Note that time can be related to redshift according to 
\begin{equation}
\label{eq:tz}
t(z)  = - \int^{\infty}_z \frac{1}{H(z')(1+z') dz'}  \, .
\end{equation}

\section{Results}
\subsection{Superheavy Dark Matter Decays}
\label{sec:DM}

In this section we consider the decays of superheavy particles with lifetimes that significantly exceed the age of the universe, $\tau_\chi \gg t_{\text{age}} \approx 4.4 \times 10^{17} \, \text{s}$, allowing them to constitute the dark matter. For simplicity, we will assume that these particles make up all the dark matter and that they decay to a single primary channel.


The neutrino flux that reaches Earth in this case includes a component from dark matter decays taking place in the halo of the Milky Way, $\Phi_{\rm G}$, as well as an isotropic flux from extragalactic dark matter decays, $\Phi_{\text{EG}}$. The Galactic component is dominated by neutrinos that were produced promptly in the superheavy particle decays. In contrast, the extragalactic flux includes significant contributions from both prompt decay products, $\Phi_{\rm EG}^{\rm prompt}$, and from neutrinos that were generated through muon or pion pair production in the resulting cascade, $\Phi^{\rm cas}_{\rm EG}$, as described in Section~\ref{sec:muons}.

%


Following Ref.~\cite{Das:2024bed}, the Galactic flux integrated over the entire sky is given by\footnote{Our results for the Galactic flux and the prompt extragalactic flux are larger than those presented in Ref.~\cite{Das:2024bed} by a factor of two. Our results sum over flavor, and over neutrinos and antineutrinos.} 
%
\begin{align}
\label{eq:fullG}
\Phi_{\rm G}(E_{\nu}) 
&= \frac{dN_{\nu}}{dE_{\nu}}(E_{\nu}) \,
\frac{1}{4 \pi m_{\chi} \tau_\chi } \int d\Omega \int_{los}
\rho_{\chi}(s,\Omega) \,ds \\
&=\frac{dN_{\nu}}{dE_{\nu}}(E_{\nu}) \, \frac{1}{4 \pi m_{\chi} \tau_\chi } \int_0^{\pi} \frac{\sin \theta}{2} \, d\theta  \, \int_0^{s_{\text{max}} (\theta)} \rho_{ \chi } (R(s)) \, ds \, , \nonumber
\end{align}
where the integrals in the top line are over solid angle, $\Omega$, and along the line-of-sight, $s$. Here, $dN_{\nu}/dE_{\nu}$ is the prompt spectrum per dark matter decay as featured in the left panels of Figures~\ref{fig:promptspectra} and~\ref{fig:promptspectrabb}. The quantity, $\theta$, is the angle between the line-of-sight and the direction of the Galactic Center, and $R(s)$ is the distance from the Galactic Center at a point along that line-of-sight:
\begin{equation}
R(s) = \sqrt{R^2_{\odot} -2 \cos{\theta} R_{\odot} +s^2},
\end{equation}
where $R_{\odot} \approx 8.34 \, \text{kpc}$ is the distance between the Galactic Center and the Solar System. We integrate the line-of-sight out to a distance of
\begin{equation}
s_{\text{max}}(\theta) = R_{\odot} \cos({\theta}) + \sqrt{R^2_h -R^2_{\odot} \sin{\theta} }  , 
\end{equation}
where we take $R_h = 100 \, \text{kpc}$ to be the size of the Galactic Halo. For the distribution of dark matter, we adopt a Navarro–Frenk–White (NFW) profile \cite{Navarro:1996gj}:
\begin{equation}
\rho_{\chi} (R) = \frac{\rho_0}{(R/R_s) \left[1 + (R/R_s) \right]^2}  \, ,
\end{equation}
with a scale radius of $R_s = 11 \, \text{kpc}$ and normalized such that the dark matter density in the solar neighborhood is $\rho_{\chi}(R_{\odot}) = 0.43 \, \text{GeV}/\text{cm}^3$. 

\begin{figure}
    \centering
    \includegraphics[width=0.49\textwidth]{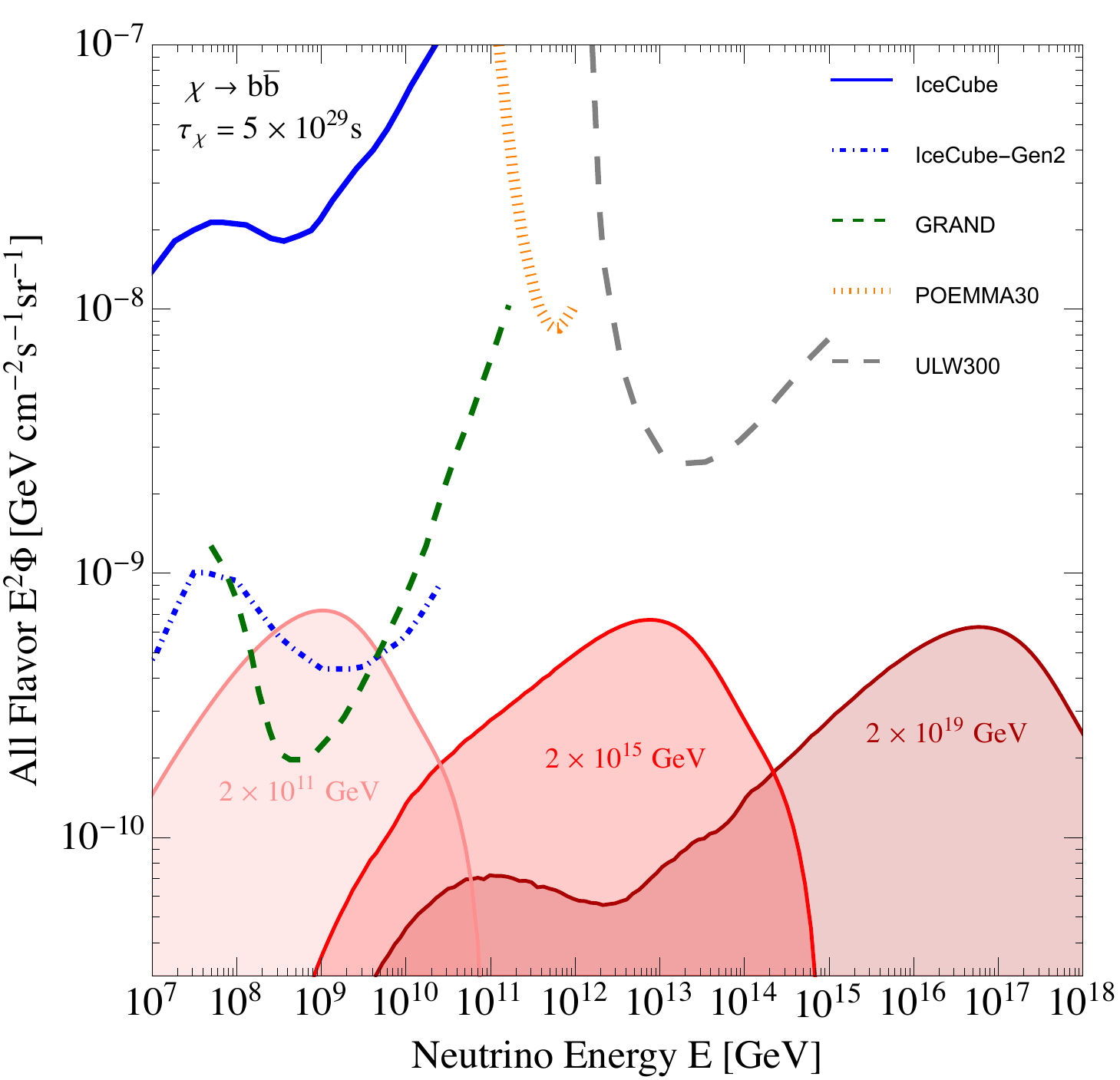}
    \includegraphics[width=0.49\textwidth]{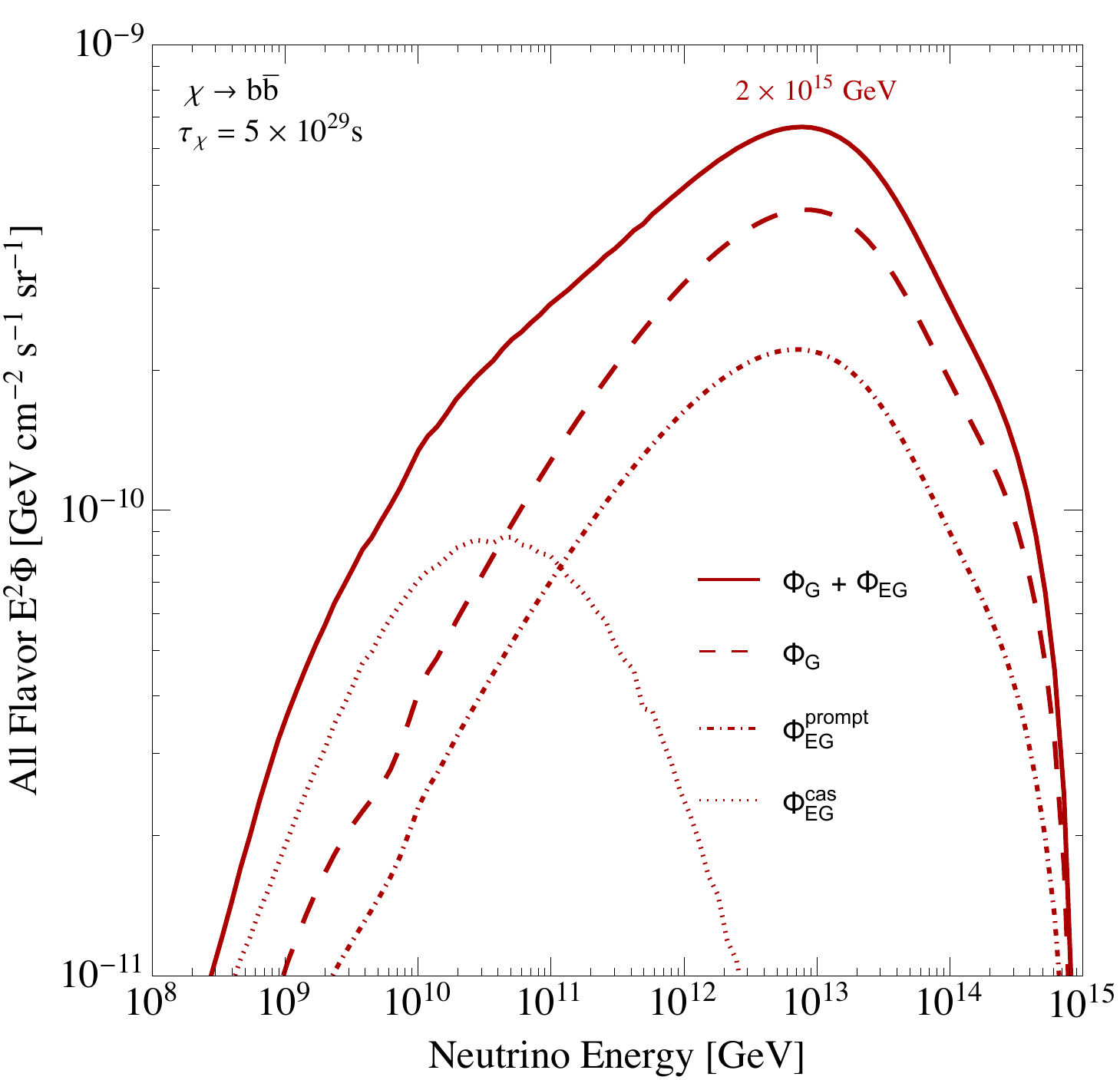}
    \caption{The all-flavor, sky-averaged neutrino flux generated by a dark matter particle with a lifetime of $\tau_{\chi} = 5 \times 10^{29} \, {\rm s}$ that decays to $\chi \rightarrow b\bar{b}$, for three values of its mass. In the left frame, this is compared to the existing constraints from the IceCube Neutrino Observatory~\cite{IceCube:2016zyt}, and to the projected sensitivities of several proposed neutrino telescopes, including IceCube-Gen2~\cite{IceCube-Gen2:2020qha}, GRAND~\cite{GRAND:2018iaj}, POEMMA30~\cite{POEMMA:2020ykm}, and ULW 300~\cite{Chen:2023mau, Das:2024bed}. 
    In the right frame, we show the neutrino spectrum for the case of $m_{\chi}=2 \times 10^{15} \, {\rm GeV}$, broken down into the contributions from decays in the Milky Way ($\Phi_{\rm G}$), from extragalactic prompt emission, ($\Phi_{\rm EG}^{\rm prompt}$), and from extragalactic cascades ($\Phi_{\rm EG}^{\rm cas}$), as generated through muon, pion, and tau pair production.
    }
    \label{fig:DMlonglived}
\end{figure}

After performing the integrals, Eq.~\ref{eq:fullG} reduces to  
\begin{equation}
\label{eq:fullG2}
\Phi_{\rm G}(E_{\nu}) = \frac{dN_{\nu}}{dE_{\nu}}(E_{\nu}) \,
\frac{J}{4 \pi m_{\chi} \tau_\chi }  \, , 
\end{equation}
where, for the specified benchmark parameters, we obtain $J \approx 3.77 \times 10^{22} \, \text{GeV/cm}^2$.

The extragalactic component of the neutrino flux is given by~\cite{Blanco:2018esa} 
\begin{align}
\Phi_{\text{EG}}(E_{\nu}) = \frac{1}{4\pi} \int \frac{\rho_{\chi}(z) \, dz}{H(z) (1+z)^3 \tau_{\chi} m_{\chi}} \, \bigg(\frac{dN_{\nu}}{dE'_{\nu}}(E'_{\nu})\bigg)_{E'_{\nu}=E_{\nu}\,(1+z)},
\end{align}
where $dN_{\nu}/dE'_{\nu}$ is the spectrum of neutrinos produced per decay, including those generated promptly and through the processes of muon, pion, and tau pair production, and $\rho_\chi$ is the energy density of the decaying particle species at redshift $z$.

In the left frame of Figure~\ref{fig:DMlonglived}, we show the all-flavor, sky-averaged neutrino flux that is generated by a dark matter particle with a lifetime of $\tau_{\chi} = 5 \times 10^{29} \, {\rm s}$ and that decays to $\chi \rightarrow b\bar{b}$, for three values of its mass. These spectra are compared to the existing constraints from the IceCube Neutrino Observatory~\cite{IceCube:2016zyt}, and to the projected sensitivities of several proposed neutrino telescopes, including IceCube-Gen2~\cite{IceCube-Gen2:2020qha} and GRAND~\cite{GRAND:2018iaj}, each after 10 years of exposure, and of POEMMA30~\cite{POEMMA:2020ykm} after 5 years of observation.
These projections are summarized in Ref.~\cite{Ackermann:2022rqc}. 
Also shown is the projected sensitivity of the Lunar Ultra-Long Wavelength radio telescope, ULW 300~\cite{Chen:2023mau, Das:2024bed}, after one year of observation.



To forecast the sensitivity of IceCube-Gen2 to decaying superheavy dark matter, we assume zero background and require the total number of observed ultra-high-energy events to exceed $N_{\nu} > 3$ in an exposure time of ten years. We calculate $N_{\nu}$ as follows:
\begin{equation}
N_{\nu} = 4 \pi  T \int_{E_{\text{min}}}^{E_{\text{max}}} \Phi(E_{\nu}) A_{\text{eff}}(E_{\nu})\, dE_{\nu}\, ,
\end{equation}
where $T=10 \, {\rm yr}$ is the exposure time, $E_{\text{min}} = 10^8 \, \text{GeV}$, $E_{\text{max}} = 10^{15} \, \text{GeV}$, and we have estimated the effective area, $A_{\text{eff}}(E_{\nu})$, from the IceCube-Gen2 sensitivity curve following the procedure described in Eq.~18 of Ref.~\cite{Das:2024bed}. We have also extrapolated the projected sensitivity of IceCube-Gen2 up to $10^{15} \, \text{GeV}$ by scaling the effective area according to the energy dependence of the neutrino-nucleon cross section~\cite{Gandhi:1995tf}. In the right frame of this Figure, we show the different contributions to the total neutrino flux.

 \begin{figure}
    \centering
    \includegraphics[width=0.322\textwidth]{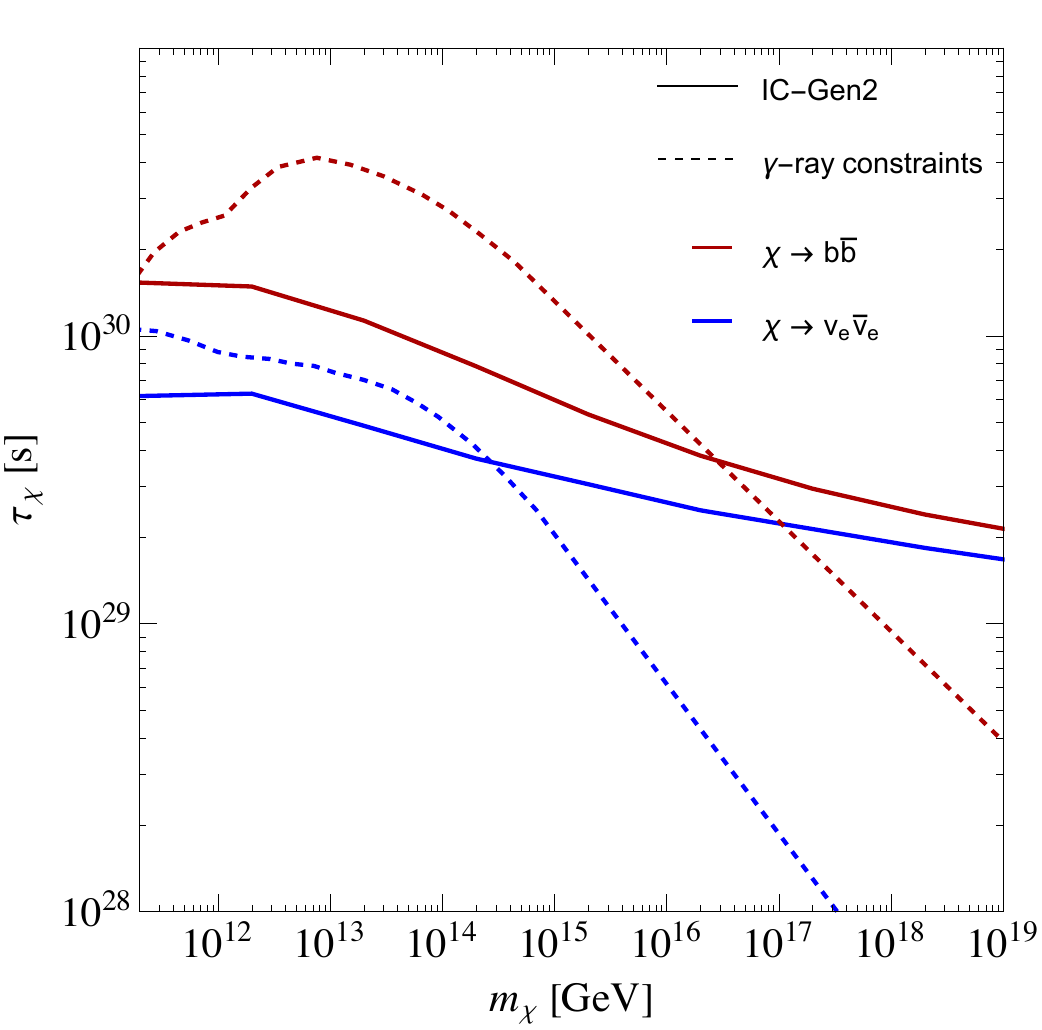}
    \includegraphics[width=0.32\textwidth]{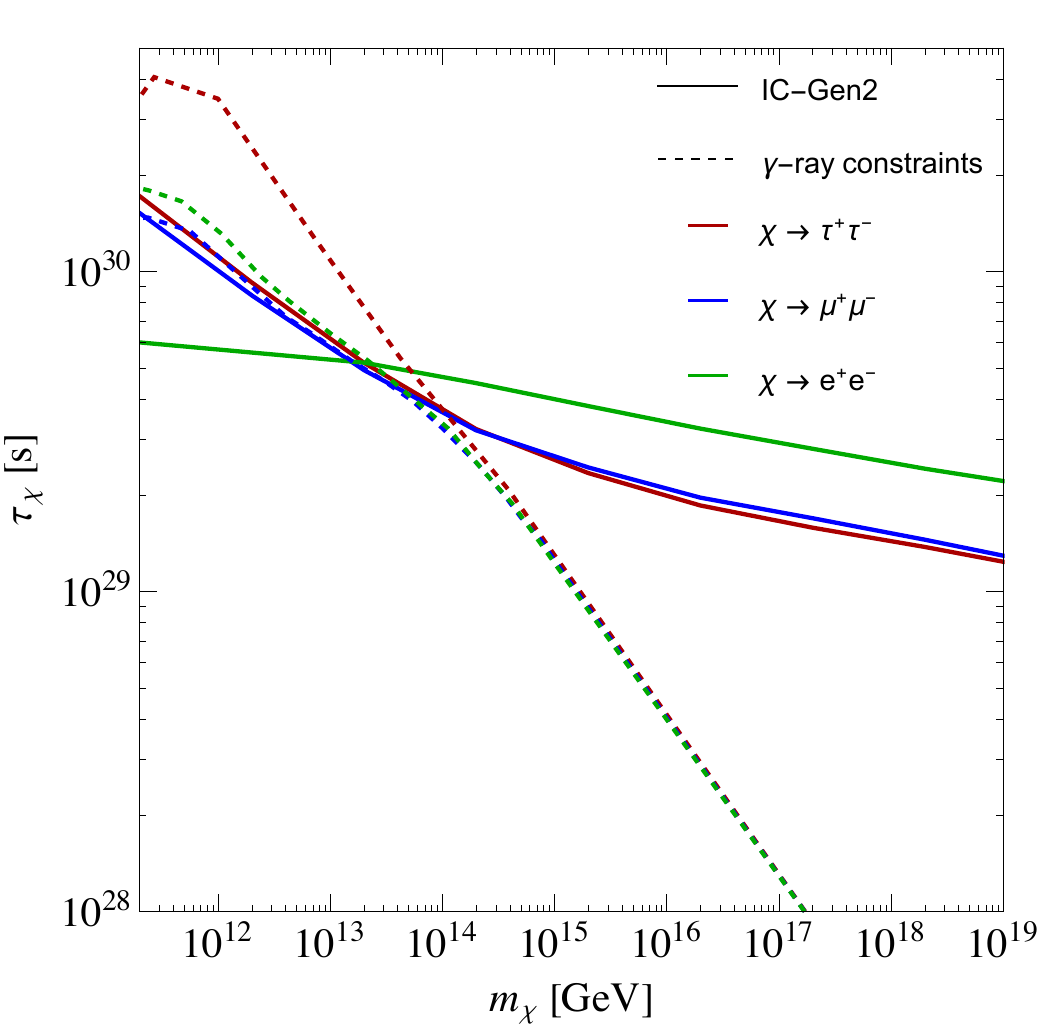}
    \includegraphics[width=0.32\textwidth]{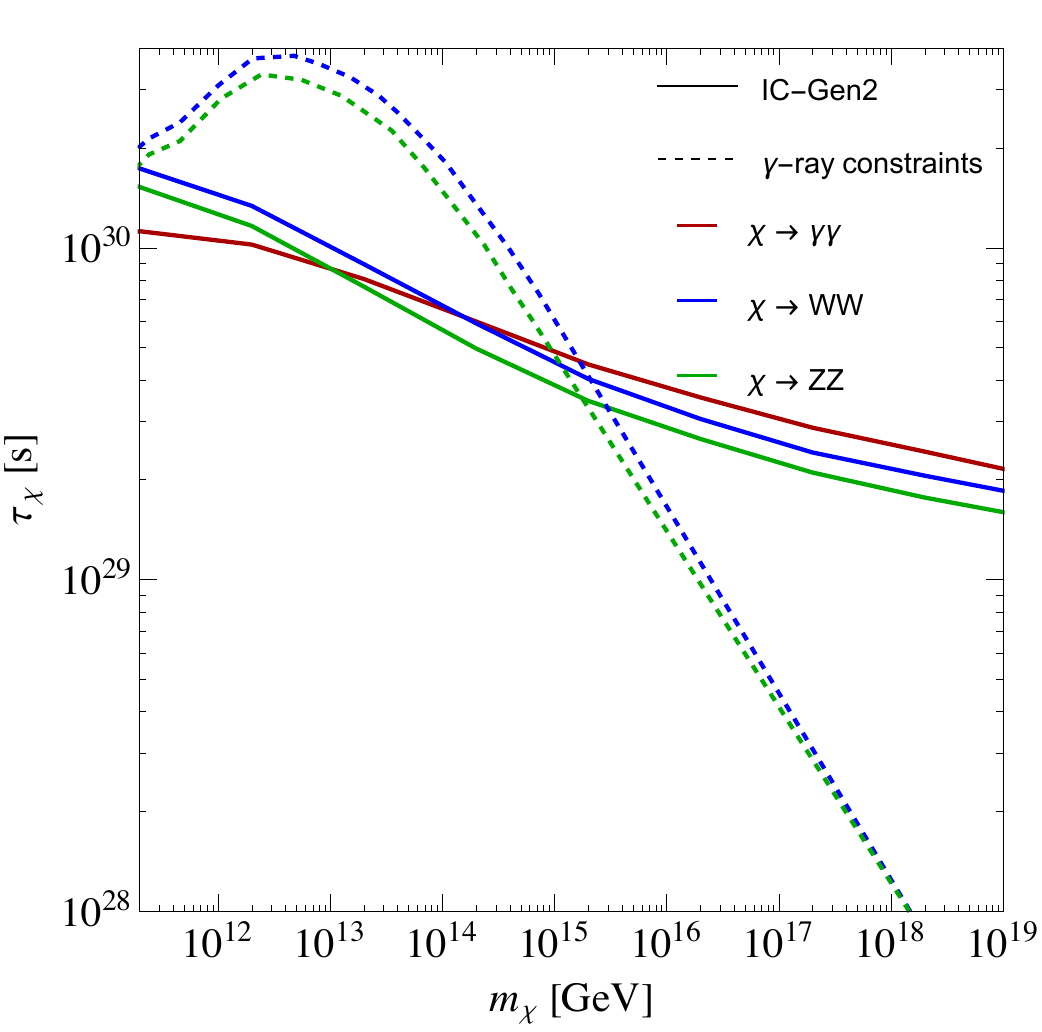}\\
 \includegraphics[width=0.32\textwidth]{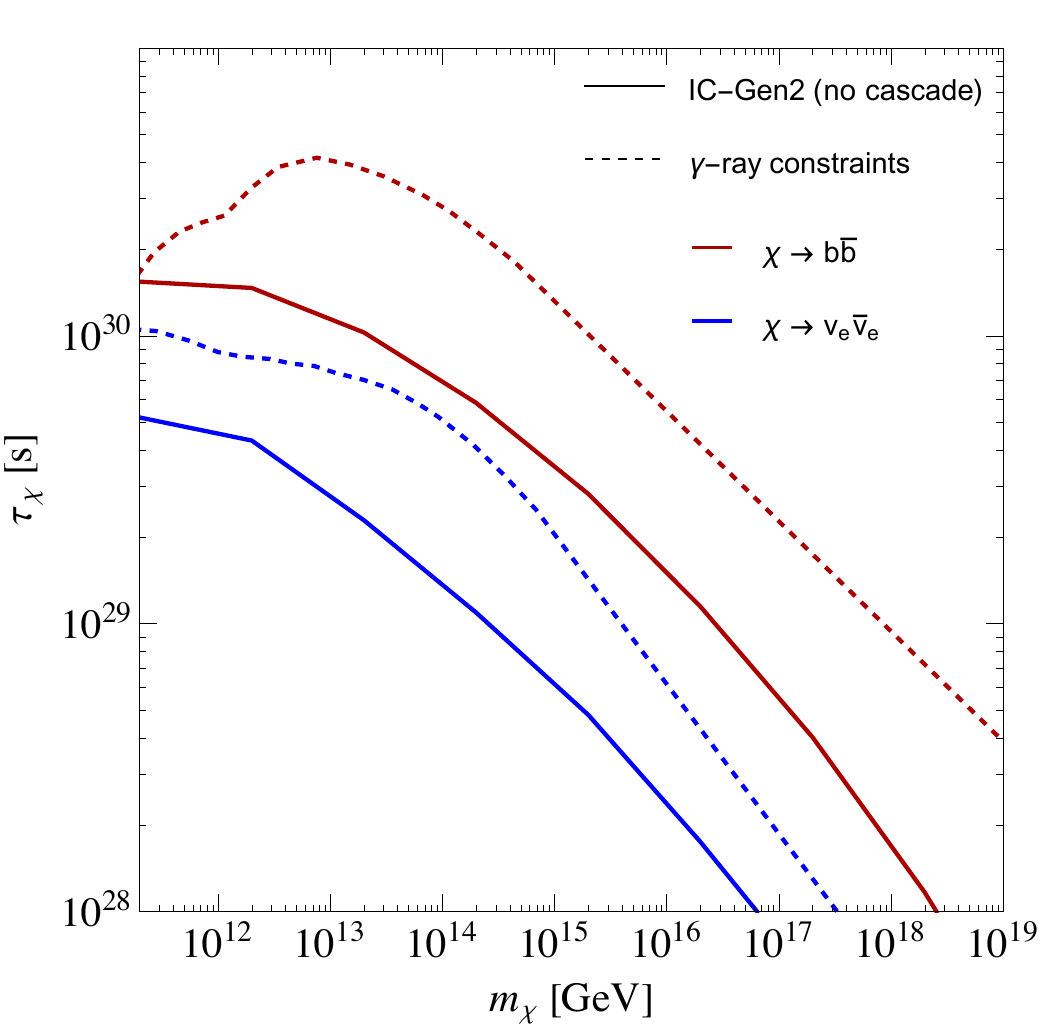}
    \includegraphics[width=0.322\textwidth]{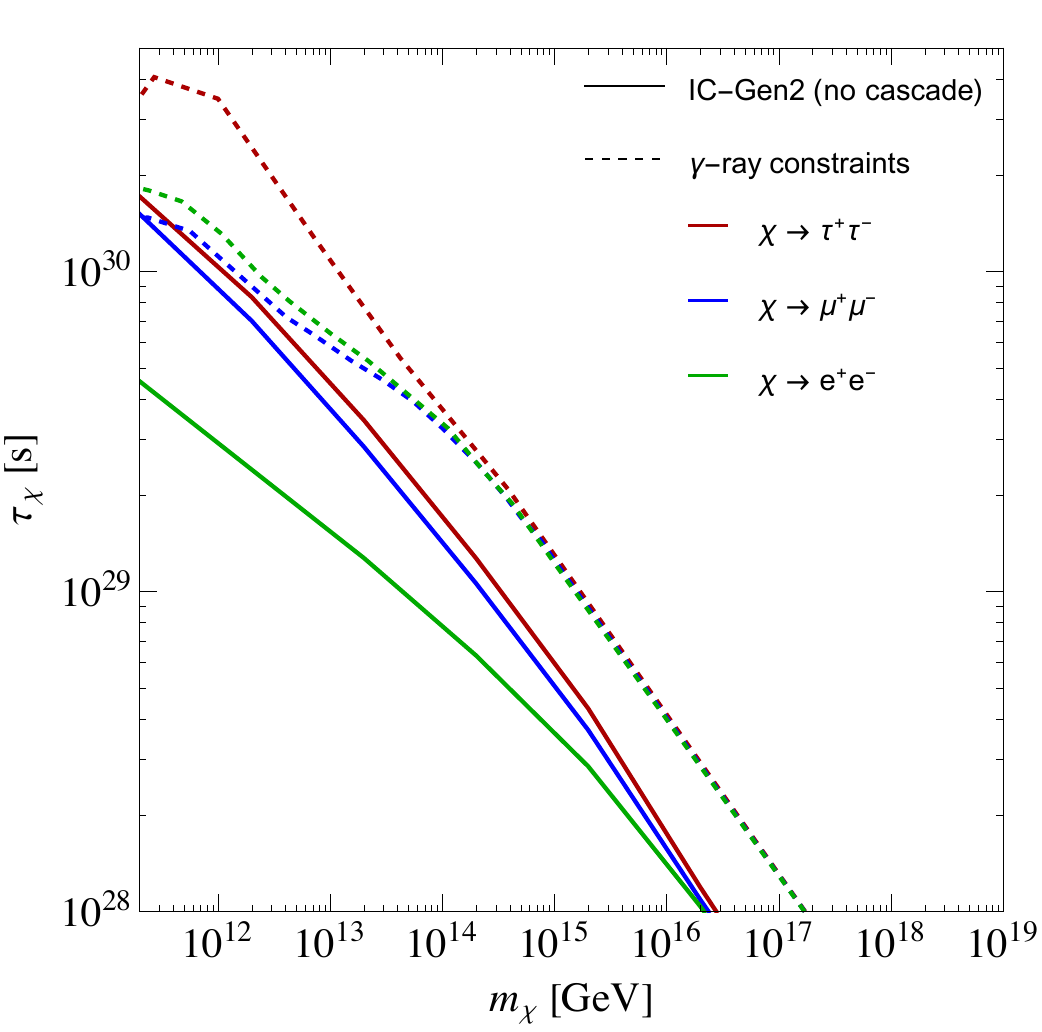}
    \includegraphics[width=0.32\textwidth]{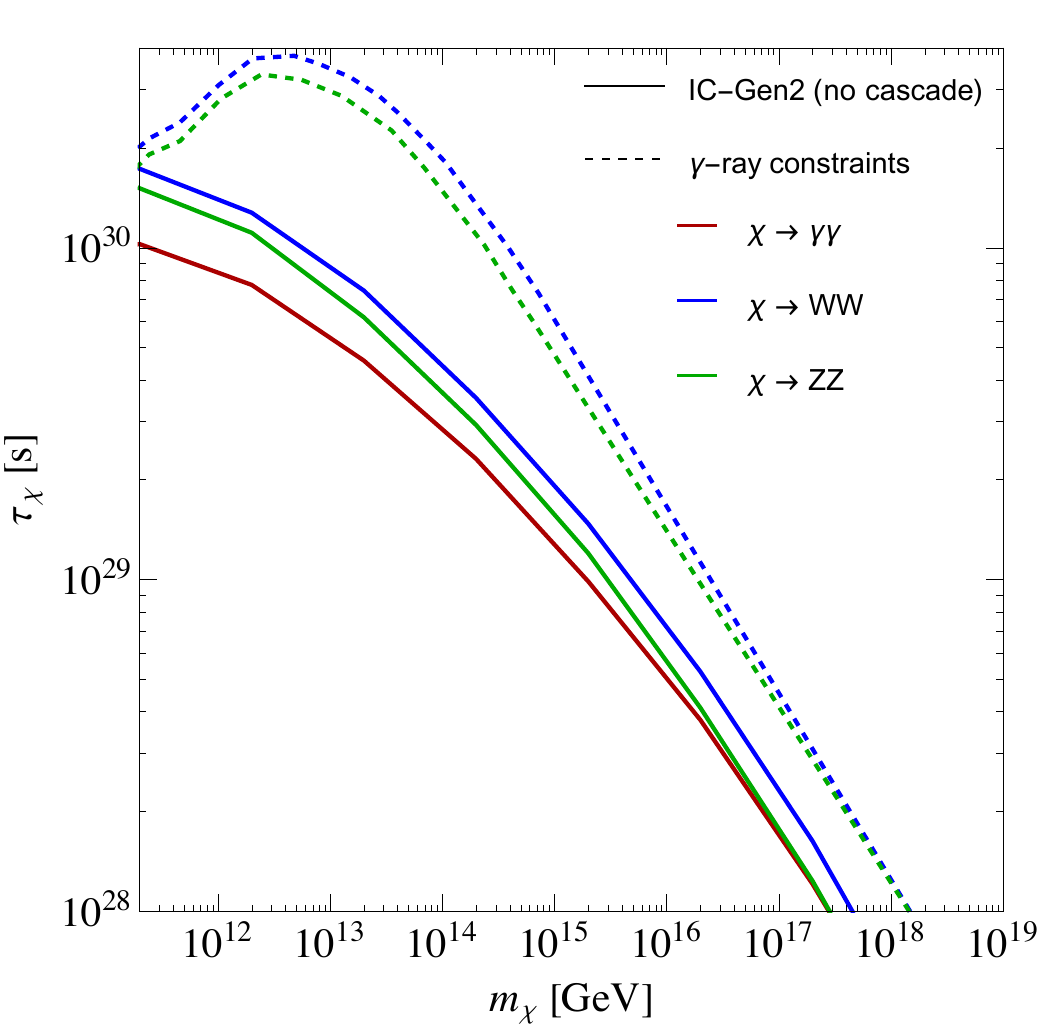}    
    \caption{The projected sensitivity of IceCube-Gen2 to decaying superheavy dark matter particles, as a function of the dark matter's mass. In the upper frames, these projected constraints are compared to the constraints derived from Galactic gamma-ray observations, as described in Ref.~\cite{Das:2023wtk}. From this comparison, we see that the projected neutrino constraints are only competitive with gamma-ray constraints for very large masses. In the lower frames, we compare the projected constraints for IceCube-Gen2 to the constraints which we would have obtained if we had neglected the processes of muon, pion, and tau pair production. For very massive dark matter particles, these processes enhance the resulting neutrino flux by more than an order of magnitude and greatly increase the prospects for detection by future ultra-high-energy neutrino telescopes.}
    \label{fig:fderfunc}
\end{figure}
%


 In Figure~\ref{fig:fderfunc}, we plot the projected sensitivity of IceCube-Gen2 to decaying superheavy dark matter particles, as a function of mass. In the upper frames, these projected constraints are compared to the constraints derived from galactic gamma-ray observations by Auger SD \cite{Savina:2021cva,PierreAuger:2022uwd,PierreAuger:2022aty}, as presented in Ref.~\cite{Das:2023wtk} (we have extrapolated these constraints for masses above $10^{15} \, \text{GeV}$). The $\chi \rightarrow \gamma \gamma$ channel was not included in the constraints derived in Ref.~\cite{Das:2023wtk}, but we expect those results to be comparable to the cases of $\chi \rightarrow  ZZ$ and $\chi \rightarrow W^+ W^-$, due to the similarity of the prompt spectra. From this comparison, we see that the projected neutrino constraints are only competitive with gamma-ray constraints for very large masses, ranging from $m_{\chi} \gsim 2 \times 10^{13} \, {\rm GeV}$ for $\chi \rightarrow e^+ e^-$ or $\chi \rightarrow \mu^+ \mu^-$, up to $m_{\chi} \gsim 3\times 10^{16} \, {\rm GeV}$ for $\chi \rightarrow b\bar{b}$. In the lower frames, we compare the projected constraints for IceCube-Gen2 to those constraints which would have been obtained if we had neglected the processes of muon, pion, and tau pair production. For very massive dark matter particles, these processes enhance the resulting neutrino flux by more than an order of magnitude and greatly increase the prospects for detection by future ultra-high-energy neutrino telescopes.

\subsection{Superheavy Long-Lived Relic Decays}

In this section, we turn our attention to superheavy particles that are long-lived, but have a lifetime that is shorter than the age of the universe, $\tau_\chi \lsim t_{\text{age}} \approx 4.4 \times 10^{17} \, \text{s}$. In this class of scenarios, the energy density of these particles evolves as
\begin{equation}
\label{eq:rhoDM}
\rho_{\chi}(z) = \Omega_{\text{dm}} \, \rho_{\rm crit} \, f_{\chi} \, (1+z)^3 \, e^{-t(z)/\tau_{\chi}} \, ,
\end{equation}
where $\Omega_\text{dm} \approx 0.265$, $\rho_{\rm crit}\approx 4.94 \times 10^{-6} \,{\rm GeV}/{\rm cm}^3$, and $f_{\chi}$ is the fraction of the dark matter that the $\chi$ population would have constituted today if it had not decayed.

\begin{figure}
    \centering
    \includegraphics[width=0.49\textwidth]{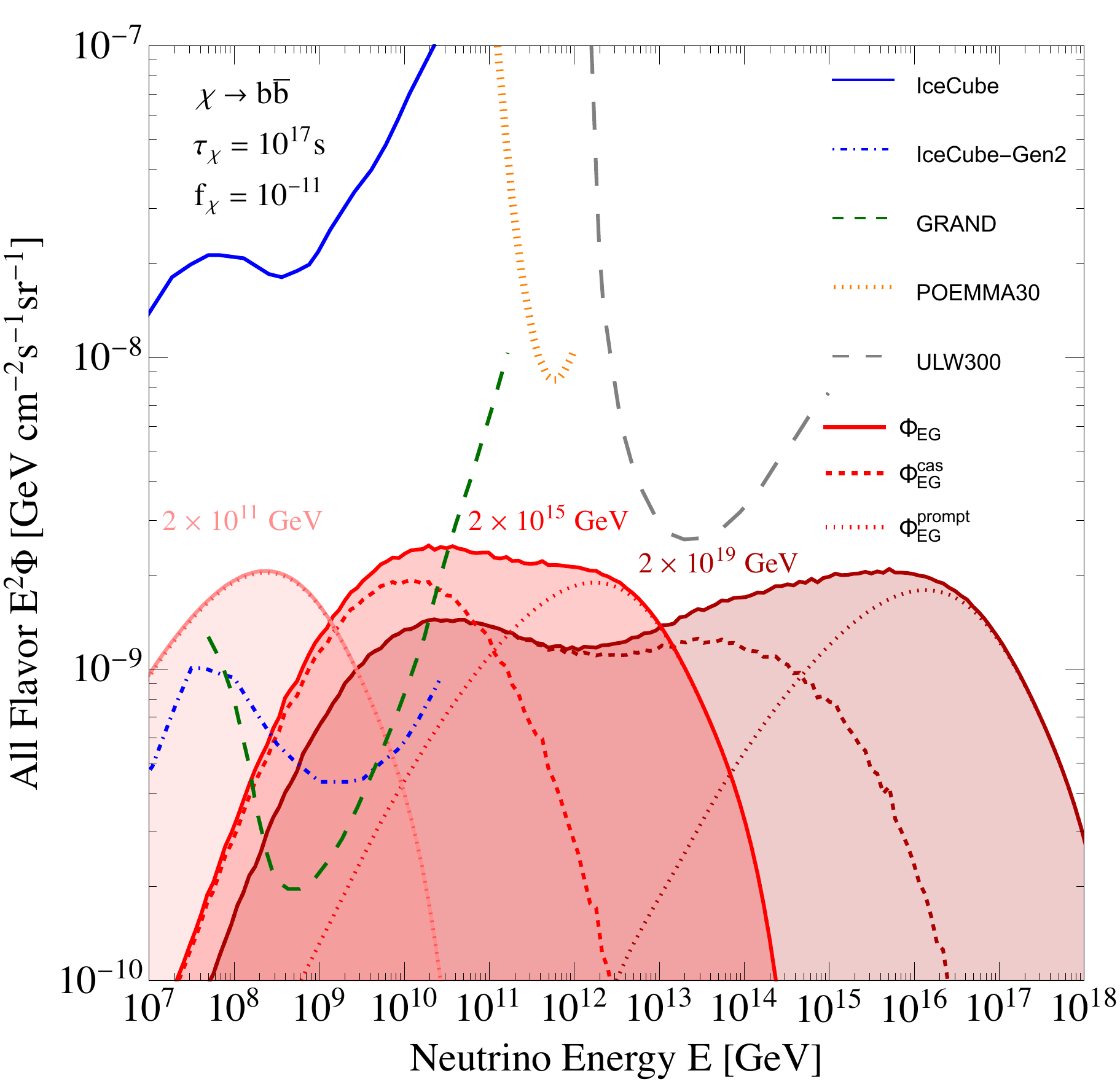}
    \includegraphics[width=0.49\textwidth]{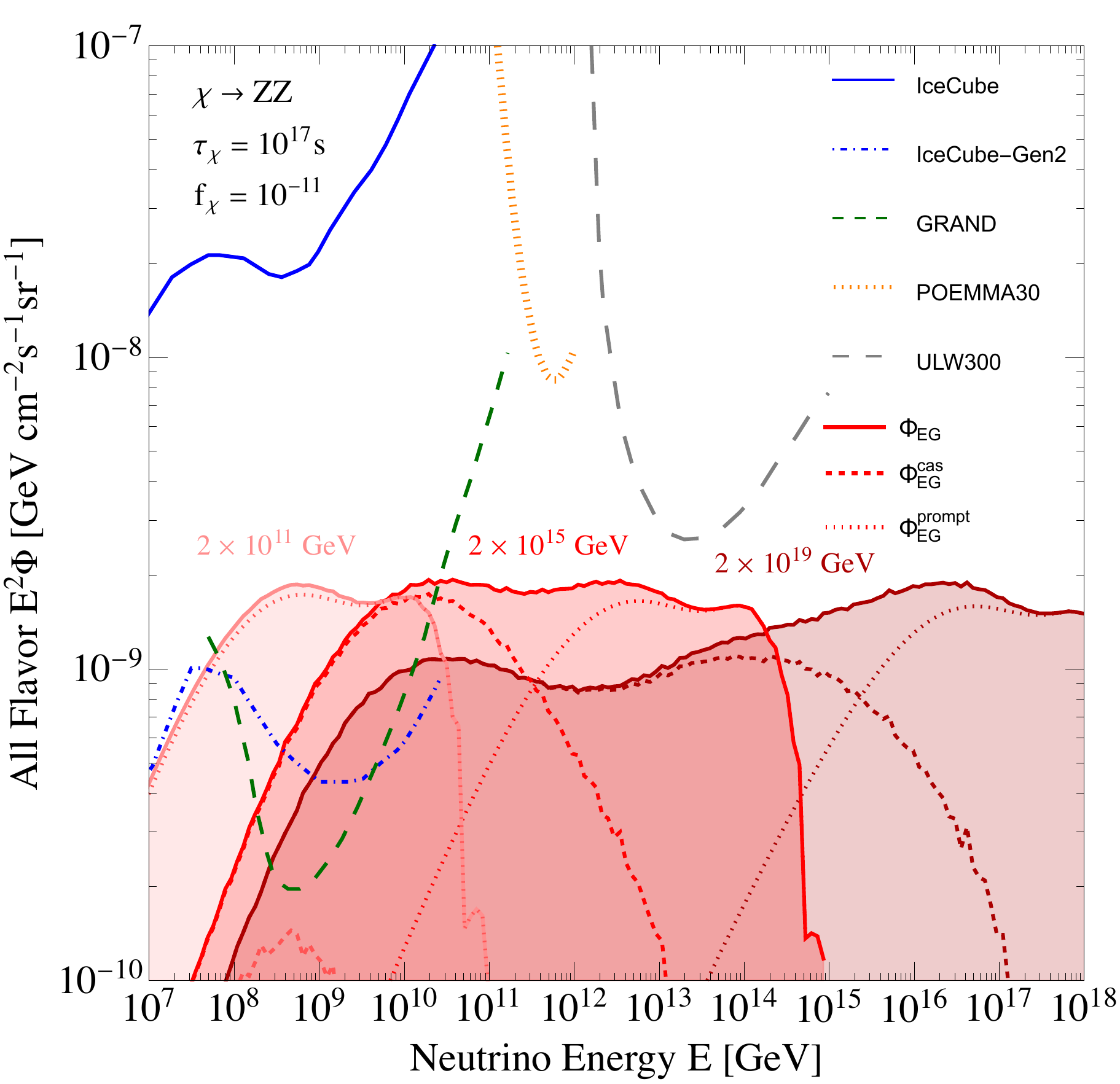}\\
        \includegraphics[width=0.49\textwidth]{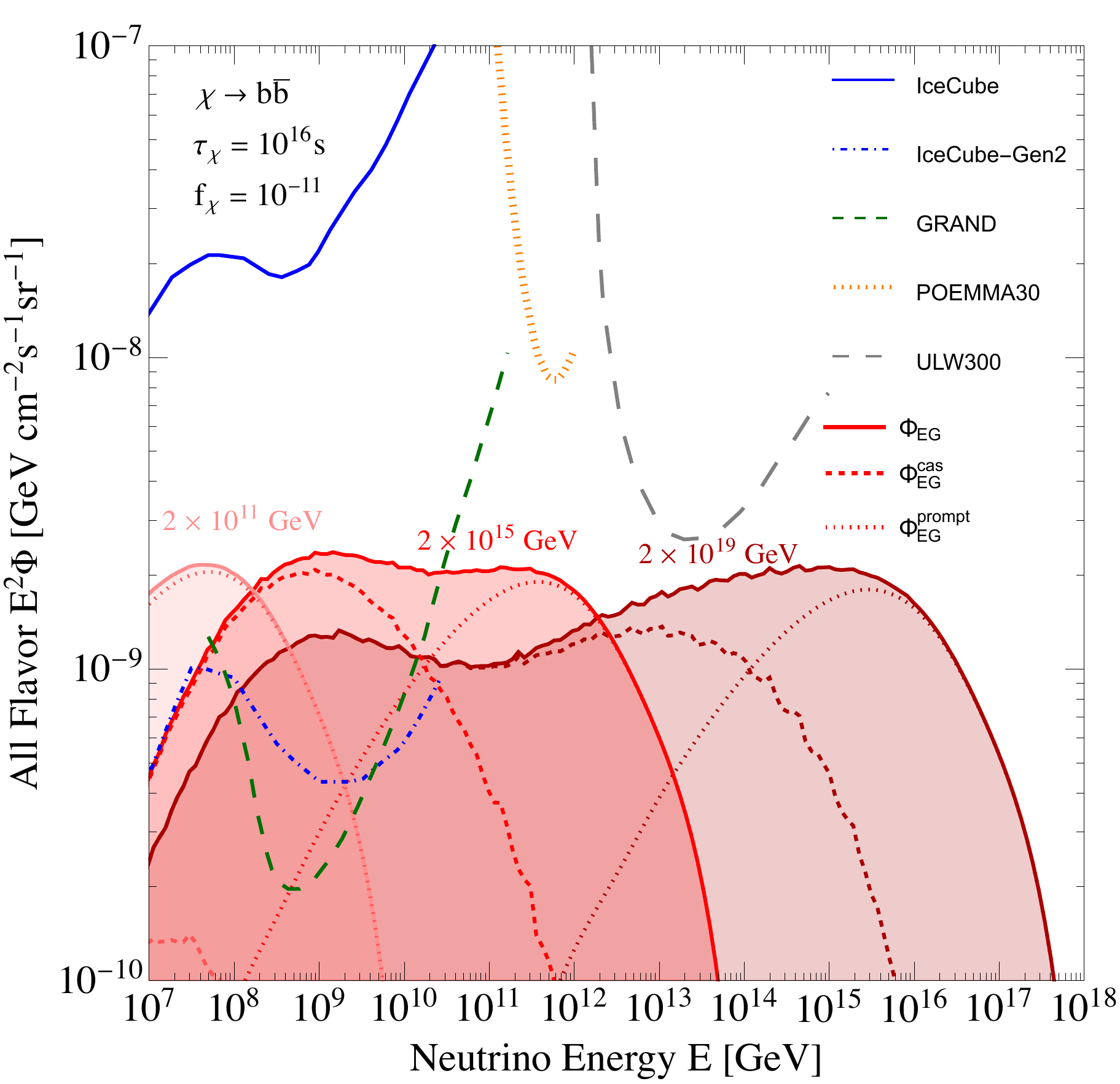}
    \includegraphics[width=0.49\textwidth]{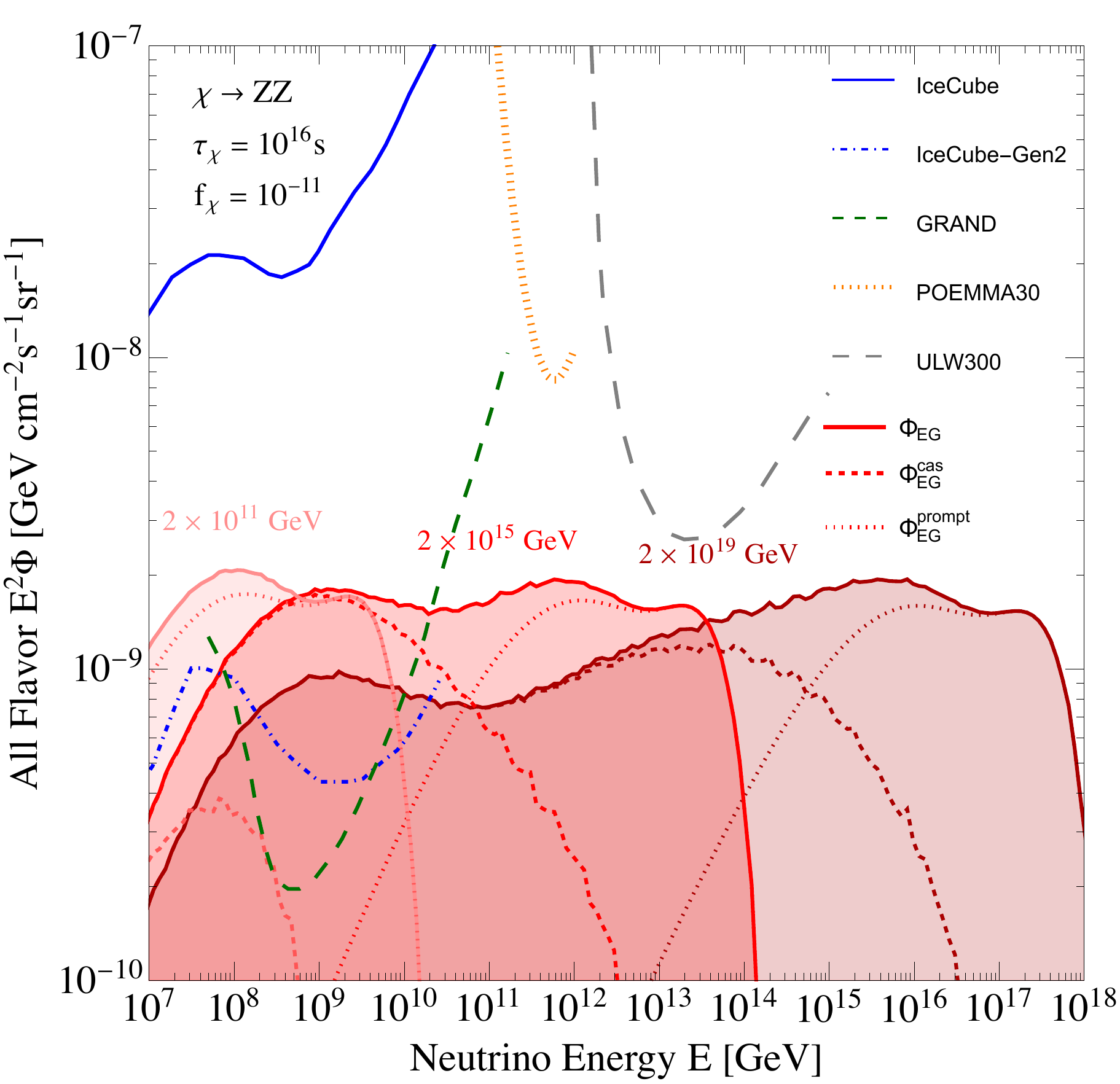}
    \caption{The all-flavor, sky-averaged neutrino flux generated by dark matter particles with a lifetime of $\tau_{\chi}=10^{17} \, {\rm s}$ (upper frames) or $\tau_{\chi}=10^{16} \, {\rm s}$ (lower frames), for the case of decays to $\chi \rightarrow b\bar{b}$ (left frames) or $\chi \rightarrow ZZ$ (right frames), and for $f_{\chi}=10^{-11}$. As in Figure~\ref{fig:DMlonglived}, we compare these neutrino fluxes to the constraints from IceCube~\cite{IceCube:2016zyt}, and to the projected sensitivity of IceCube-Gen2~\cite{IceCube-Gen2:2020qha} and other future high-energy neutrino telescopes~\cite{GRAND:2018iaj,POEMMA:2020ykm,Chen:2023mau, Das:2024bed}. These neutrino spectra are further broken down into their contributions from prompt emission, and from muon and pion pair production in the subsequent cascade.}
    \label{fig:fluxbb17}
\end{figure}

For $\tau_{\chi} \ll t_{\rm age}$, few of these particles will be left in the Galactic Halo, leaving us with a neutrino flux that is generated predominantly by extragalactic decays, taking place at relatively high redshifts ($z (10^{17} \, \text{s}) \approx 2$; $z (10^{16} \, \text{s}) \approx 13$). In this study, we will further restrict our attention to cases in which $\tau_{\chi} \geq 10^{16} \, \text{s}$, for which the majority of decays take place at $z \lsim 15$. For shorter lifetimes, neutrino scattering with the cosmic neutrino background through the $Z$ resonance can significantly impact the resulting neutrino flux. We leave the exploration of this interesting 
regime to future study.

In Figure~\ref{fig:fluxbb17}, we show the all-flavor, sky-averaged neutrino flux generated by dark matter particles with a lifetime of $\tau_{\chi}=10^{17} \, {\rm s}$ (upper frames) or $\tau_{\chi}=10^{16} \, {\rm s}$ (lower frames), for the case of decays to $\chi \rightarrow b\bar{b}$ (left frames) or $\chi \rightarrow ZZ$ (right frames), and for $f_{\chi}=10^{-11}$. As in Figure~\ref{fig:DMlonglived}, we compare these neutrino fluxes to the constraints from IceCube~\cite{IceCube:2016zyt}, and to the projected sensitivity of IceCube-Gen2~\cite{IceCube-Gen2:2020qha}, and other future high-energy neutrino telescopes~\cite{GRAND:2018iaj,POEMMA:2020ykm,Chen:2023mau, Das:2024bed}. These neutrino spectra are further broken down into their contribution from prompt emission, and from muon, pion, and tau pair production in the subsequent cascade. We can see that the contributions from the cascade dominate the spectrum below the largest possible neutrino energies.

\begin{figure}
    \centering
    \includegraphics[width=0.49\textwidth]{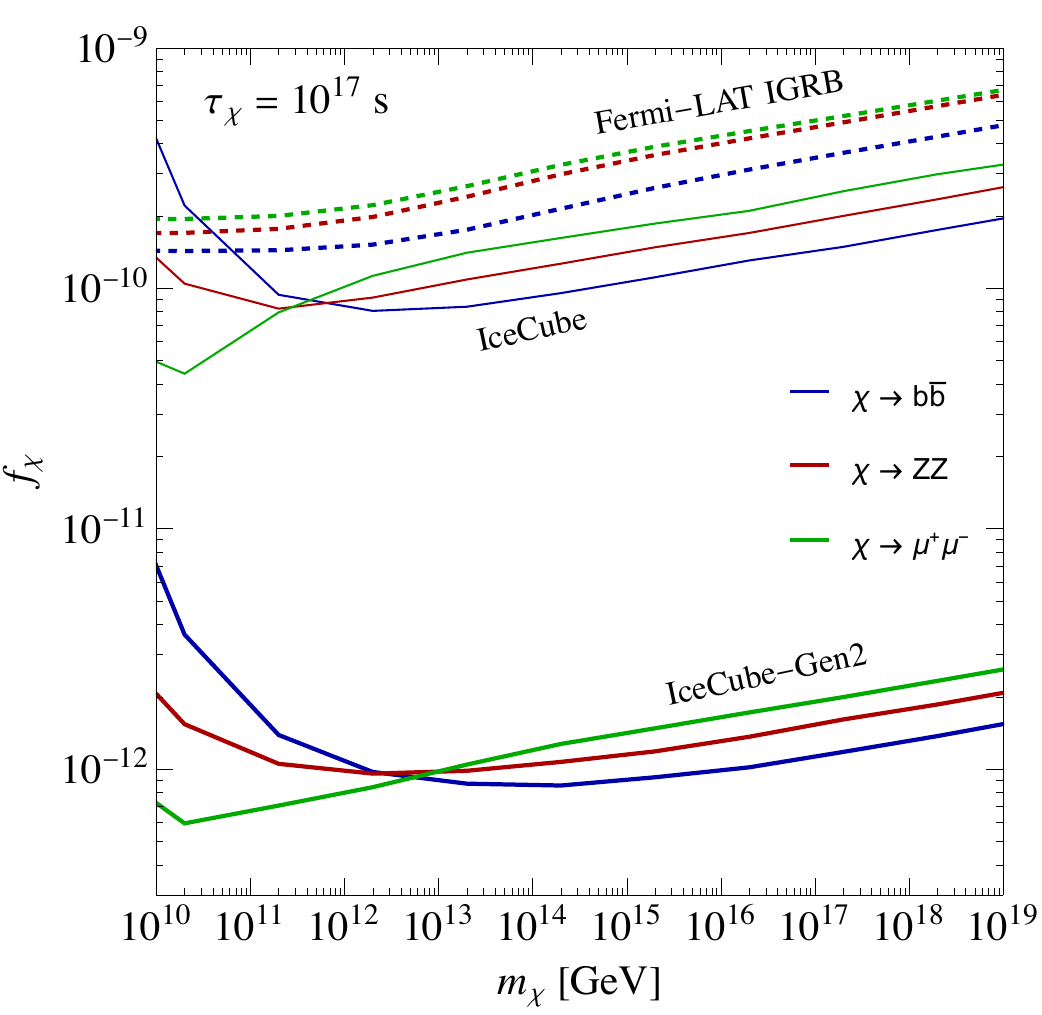}
\includegraphics[width=0.49\textwidth]{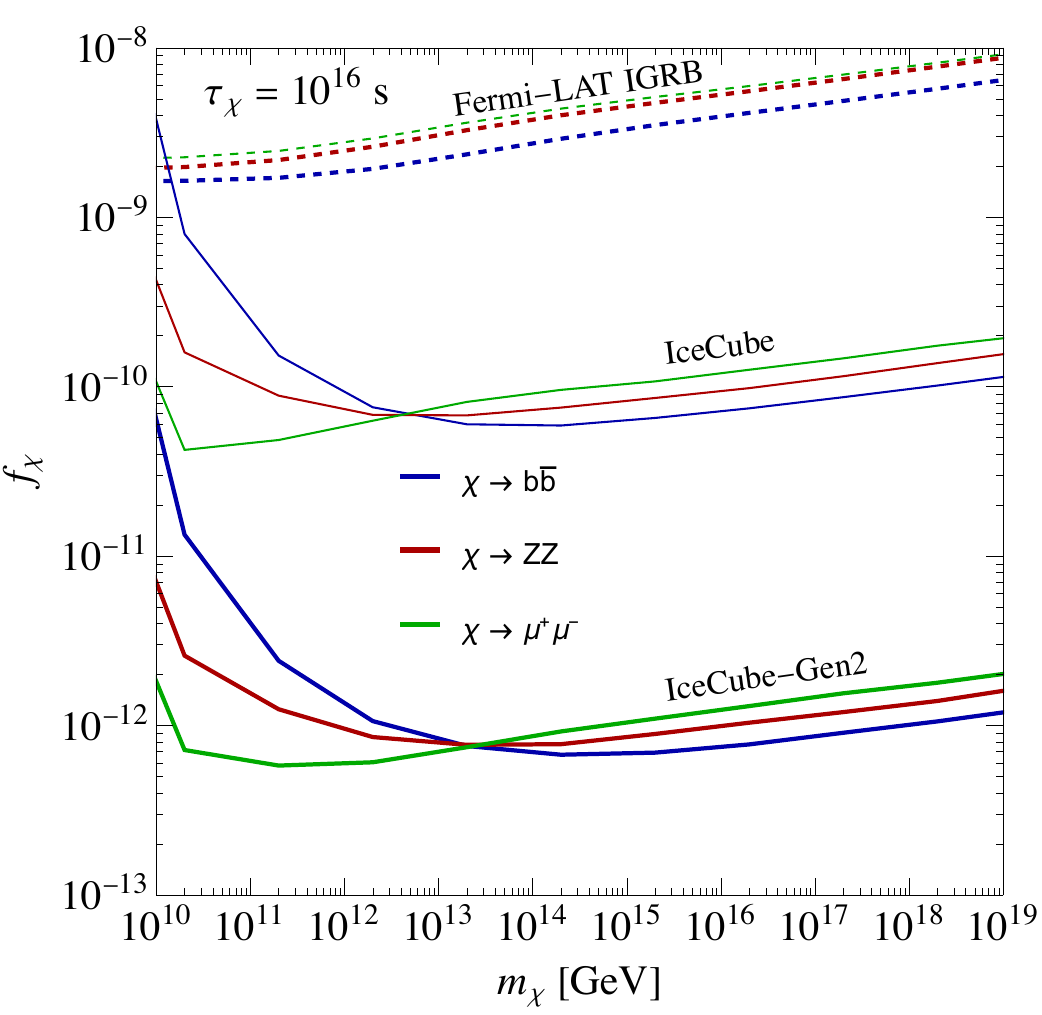} \\
  \includegraphics[width=0.49\textwidth]{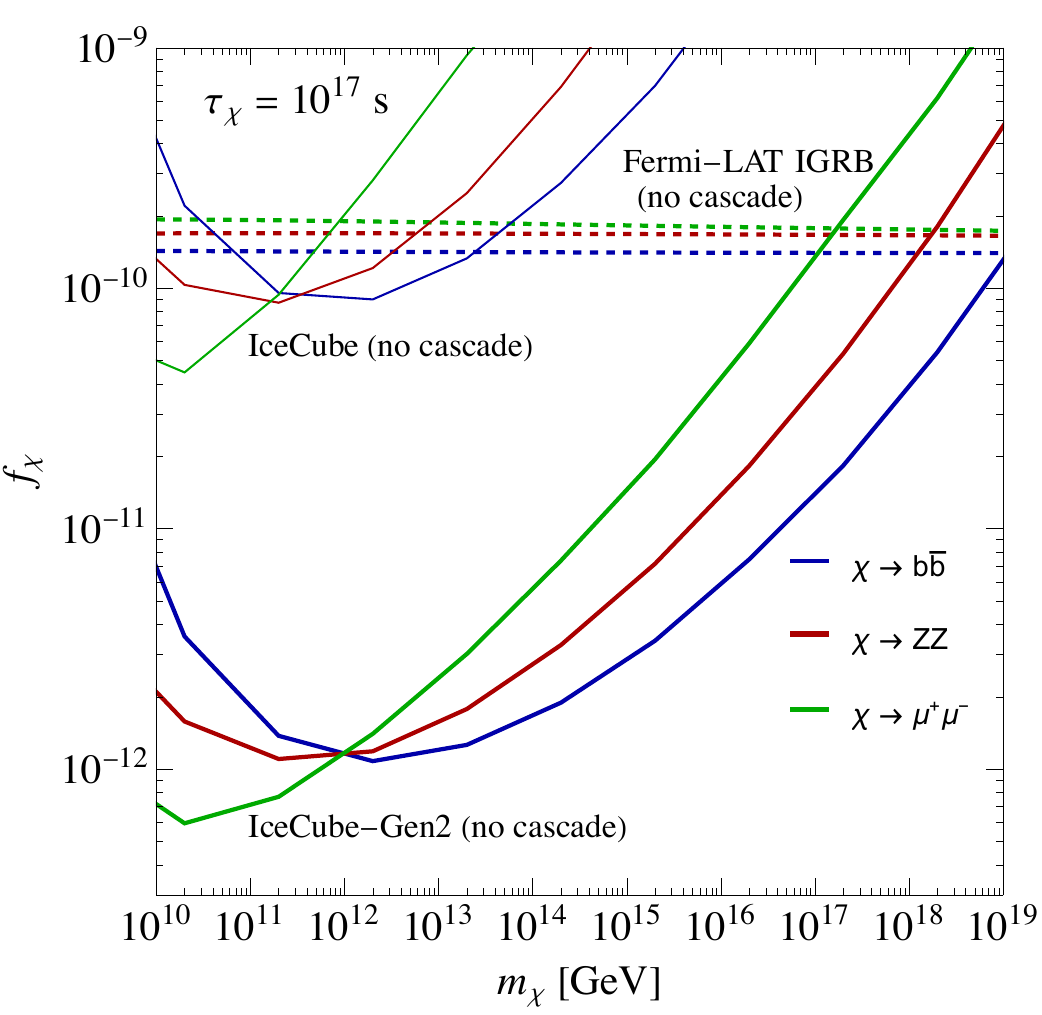}
\includegraphics[width=0.49\textwidth]{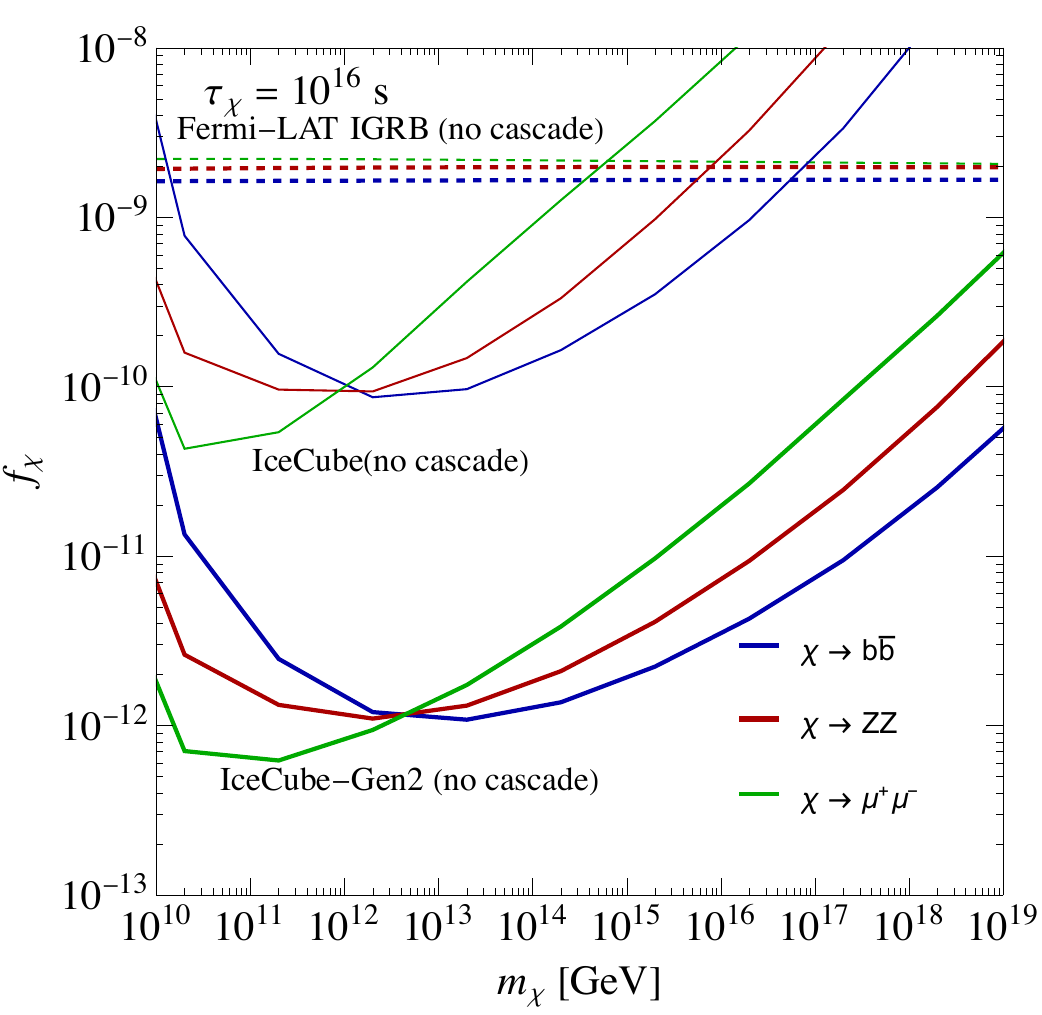}
    \caption{ 
    Constraints from IceCube, and the projected sensitivity of IceCube-Gen2, to decaying superheavy particles as a function of mass, for lifetimes of $\tau_{\chi}=10^{16} \, {\rm s}$ (left frames) or $\tau_{\chi}=10^{16} \, {\rm s}$ (right frames). In the upper frames, these projected constraints are compared to the constraints derived from gamma-ray observations of the isotropic gamma-ray background (IGRB). From this comparison, we see that the current constraints from IceCube already outperform those provided by Fermi in the case of $\tau_{\chi}=10^{17} \, {\rm s}$, and are more than an order of magnitude more powerful than gamma-ray constraints in the case of $\tau_{\chi}=10^{16} \, {\rm s}$. IceCube-Gen2 will further improve upon these limits by approximately two orders of magnitude. In the lower frames of this figure, we compare the projected sensitivity from IceCube, IceCube-Gen2, and Fermi LAT which would have been obtained if we had neglected the processes of muon, pion, and tau pair production. For very massive dark matter particles, these processes enhance the resulting neutrino flux by more than an order of magnitude, and greatly enhance the prospects for detection by future ultra-high-energy neutrino telescopes.}
    \label{fig:lifetimes}
\end{figure}

We now again
consider the constraints that can be derived from gamma-ray observations. In the previous subsection, in considering scenarios for which $\tau_{\chi} \gg t_{\rm age}$, the constraints were dominated by searches for ultra-high-energy photons from the halo of the Milky Way. For the case at hand, $\tau_{\chi} \lsim t_{\rm age}$, most of the superheavy particles in the Galactic Halo have already decayed, leaving us instead to be more restricted by the intensity of the GeV-TeV isotropic gamma-ray background, as measured by the Fermi-LAT~\cite{Fermi-LAT:2014ryh}. Here, we derive these gamma-ray constraints by demanding that the photon flux from the decaying relics as calculated in Eq.~\eqref{eq:gammacon} remains below the measured $95\%$ upper limit CL. We note that the highest energy bin  dominates our constraints.

Finally, in Figure~\ref{fig:lifetimes}, we show the neutrino constraints from IceCube, the gamma-ray constraints from the Fermi Gamma-Ray Space Telescope, and the projected sensitivity of IceCube-Gen2, to decaying particles, as a function of mass, for lifetimes of $\tau_{\chi}=10^{17} \, {\rm s}$ (left frames) or $\tau_{\chi}=10^{16} \, {\rm s}$ (right frames), and for selected decay channels, $\chi \rightarrow b\bar{b}$, $ZZ$, and $\mu^+\mu^-$. In the upper frames, the neutrino constraints are compared to the constraints based on gamma-ray observations of the isotropic gamma-ray background (IGRB). From this comparison, we see that the current constraints from IceCube
already outperform
those provided by Fermi LAT in the case of $\tau_{\chi}=10^{17} \, {\rm s}$, and are more than an order of magnitude more powerful than gamma-ray constraints in the case of $\tau_{\chi}=10^{16} \, {\rm s}$. IceCube-Gen2 will further improve upon these limits by approximately two orders of magnitude. In the lower frames of Figure~\ref{fig:lifetimes}, 
we show the comparison of constraints set by IceCube, and projected constraints by IceCube-Gen2 to the gamma-ray constraints from the Fermi LAT if one had neglected the cascades from muon, pion, and tau pair production. For very massive dark matter particles, these processes enhance the resulting neutrino flux by more than an order of magnitude, and simultaneously reduce the predicted gamma-ray flux, thereby greatly increasing the prospects for detection with future ultra-high-energy neutrino telescopes.

\section{Summary and Conclusions}

In this study, we have revisited the sensitivity of ultra-high-energy neutrino telescopes to the decays of superheavy particles. For particles heavier than $m_{\chi} \gsim 10^{10}\,  {\rm GeV}$, the photons produced in their decays can exceed the thresholds for muon and pion pair production, enhancing the predicted flux of ultra-high-energy neutrinos. Over the course of the development of the electromagnetic cascade, this process can transfer an order one fraction of the total energy of the decay products into neutrinos, relaxing the constraints on superheavy particle decays that can be derived from gamma-ray observations, and strengthening the limits from IceCube, as well as enhancing the projected sensitivity of next-generation neutrino telescopes, such as IceCube-Gen2.

For particles with lifetimes greater than the age of the universe, the sensitivity of IceCube-Gen2 is projected to exceed the constraints from gamma-ray observations only in the case of very large masses, $m_{\chi} \gsim 10^{13}-10^{16} \, {\rm GeV}$, depending on the primary decay channel. For somewhat shorter-lived particles, $\tau_{\chi} \sim 10^{16}-10^{17} \, {\rm s}$, we find that the most stringent constraints come from neutrino observatories. In particular, IceCube provides the strongest constraints on this class of scenarios for all $m_{\chi} \gsim 10^{10} \, {\rm GeV}$. We anticipate that IceCube-Gen2 will improve upon IceCube's sensitivity to superheavy particle decays by approximately two orders of magnitude.

\bigskip

\acknowledgments

We would like to thank Nick Rodd and Rocky Kolb for helpful discussions. KB thanks the U.S. Department of Energy, Office of Science, Office of High Energy Physics, under Award Number DE-SC0011632 and the Walter Burke Institute for Theoretical Physics. DH is supported by the Office of the Vice Chancellor for Research at the University of Wisconsin-Madison, with funding from the Wisconsin Alumni Research Foundation. ES is supported by the U.S. National Science
Foundation Graduate Research Fellowship Program, Grant
Number 2140001.



\bibliographystyle{JHEP}
\bibliography{wimpzilla}


\end{document}